\documentclass{IEEEtran}
\usepackage{natbib}

\usepackage{amsmath}
\usepackage{amssymb}
\usepackage{graphicx}
\usepackage{physics}

\usepackage{caption}
\usepackage[bf]{subfigure}

\makeatletter
\renewcommand{\@thesubfigure}{\alph{subfigure})}
\makeatother

\begin{document}

\title{Traveltime signature of 3-D edge diffractions exemplified by triple-square-root moveout
}

%\address{
%Institute of Geophysics, 
%University of Hamburg, 
%Bundesstra{\ss}e 55,  
%Hamburg, 20146 Germany
%}

%pavel.znak@uni-hamburg.de
%dirk.gajewski@uni-hamburg.de \\

\date{}
%\author{Pavel Znak and Dirk Gajewski}

\author{
    \IEEEauthorblockN{Pavel Znak\IEEEauthorrefmark{2} and Dirk Gajewski\IEEEauthorrefmark{2}}\\
    \IEEEauthorblockA{\IEEEauthorrefmark{2}Institute of Geophysics, University of Hamburg, Bundesstra{\ss}e 55, Hamburg, 20146 Germany}
}

%\lefthead{Znak \& Gajewski}
%\righthead{Traveltime signature of 3-D edge diffractions}
\maketitle

\begin{abstract}
Various underground anomalies, both natural and artificial, cause diffraction of high-frequency seismic and electromagnetic pulses emitted from the earth's surface. 
Backscattered, they are registered by seismic sensors and ground-penetrating radars. Most of these signals can be categorized as either point or edge diffraction.
Despite the abundance of linear structures in geological formations and among buried anthropogenic objects, diffraction processing often relies on the idea of point diffraction. However, 3-D edge diffractions have unique properties that need to be exploited. We show that the mixed source-receiver traveltime derivatives, available from data, identify edge diffractions at arbitrary offsets.
Additionally, they constitute a system of ordinary differential equations describing finite-offset focusing curves on the acquisition surface. Such curves enable sorting of the data into fragments that focus on specific points on the edge. To confirm and demonstrate these properties, we derive an exact traveltime formula for a straight diffractor in a homogeneous medium. We refer to it as the triple-square-root moveout in analogy to the double-square-root moveout of 2-D diffractions. Since it represents a potentially useful approximation, we compare it with standard moveouts and wave modeling.
\end{abstract}

\section{Introduction}
Diffracted waves carry information about the underground complementary to that brought by reflections. 
The most frequently observed in applied seismic and ground-penetrating radar records can be qualified as either point or edge diffractions. 
Small cavities, inclusions, or structural tips cause point diffraction. Edge diffractions arise on fractures, faults, unconformities, pinch-outs, etc. There are also numerous anthropogenic scatterers with kinematics of edge diffraction, such as tunnels, pipes, and communication cables. 
Note that this classification is based on kinematics, ignoring particular configurations that form the diffractors. This way, it makes no distinction between edges of surfaces, edges of wedges, or thin linear objects. The very concept of diffractor, and therefore such classification, is relevant for objects having at least one radius of curvature small enough in comparison to the predominant wavelength. 

\citet{Sommerfeld1896} was the first to solve the problem of electromagnetic plane wave diffraction on an opaque half-plane in a homogeneous medium. He also generalized his approach to diffraction on wedges with angles that are fractions of $\pi$ \citep{Sommerfeld1912}. \citet{Macdonald1913} completed the derivations for arbitrary angles. Half a century later, \citet{Keller1962} introduced a general approach to the high-frequency asymptotics of edge diffractions in heterogeneous media, where bending diffracted rays satisfy a geometrical law on the edge, and the ray amplitudes are derived from solving local plane-wave diffraction problems of different configurations. For brevity, we refer to the geometrical law of edge diffraction as Keller's law. A second-order traveltime approximation for 3-D edge diffractions is derived by \citet{Sun1994}.

There are two main strategies for imaging subsurface diffractors. The first is to modify a depth imaging algorithm such that it emphasizes the contribution by diffractions and suppresses the one by reflections. One can achieve this through filtration based on specularity and Snell's law \citep{Kozlov2004, Moser2008, Li2020}. To emphasize linear structures in 3-D, imaging can be performed exclusively for the traces that correspond to Keller's law \citep{Alonaizi2013,Keydar2019, Rochlin2021}. \citet{Znak2022} suggest filters based on both Snell's and Keller's laws to separate 3-D images of points, edges, and reflectors. An alternative approach to modifying the imaging algorithm is to filter reflections in the common dip-angle image gathers \citep{Reshef2009, Klokov2012}.

The second main strategy involves a preliminary identification and separation of diffracted wavefields based on a distinguishing property. When diffractions are isolated, one can use a standard approach to get the diffraction image. The identification step requires no prior knowledge of the subsurface. It is valuable in its own right, as the isolated diffractions can serve other purposes, such as model building. Different principles may underlie the separation process. In this regard, we limit ourselves to two of them.

One research direction utilizes the appearance of diffractions in the zero-offset sections. They have diverse slopes in comparison to less curved reflections. Starting from this observation, \citet{Fomel2002} and further \citet{Fomel2007} formulate the plane-wave destruction filter. \citet{Merzlikin2020} isolate edge diffractions with its azimuthal extension. Such a criterion, however, may show ambiguity in
a medium with substantial lateral variations. The curvature of reflections in shot gathers complicates its application for finite offsets. Another challenge is to separate the three wave types since the 3-D edge diffraction behaves as either reflection or diffraction, depending on the azimuth.

Another approach, to which this paper belongs, exploits exact identities for second-order traveltime derivatives and, equivalently, for wavefront curvatures. The wavefront attributes derived from data by automatic coherence analysis provide estimates of these quantities. In 2-D, there are scalar identities for the zero-  and finite-offset wavefront attributes, enabling the identification of diffractions in the stack \citep{Jaeger2001} and pre-stack \citep{Zhang2001} domains. There is a limitation caused by caustics, which do not allow for the second-order traveltime derivatives. Apart from that, these identities hold for any model, including heterogeneity and anisotropy. 
They unambiguously distinguish diffractions from reflections. In practice, with band-limited and strongly interfering signals, however, it comes down to a technical problem of extracting the wavefront attributes such that they accurately represent the traveltime derivatives.
\citet{Dell2011} apply the zero-offset criterion in the form of a smoothed threshold to overcome their inaccuracy. \citet{Rad2018} extend this workflow to 3-D based on an identity of the zero-offset wavefront curvature matrices, which is valid for point diffractions. 
\citet{Schwarz2019} combines these criteria with the one based on the slope diversity to isolate reflections and adaptively subtract them from the total wavefield, resulting in diffraction sections.
\citet{Znak2023} formulate a general criterion in terms of the zero-offset wavefront curvature matrices to unambiguously discriminate all three wave types in 3-D: point diffractions, edge diffractions, and reflections. The authors also introduced an attribute-based system of ordinary differential equations to compute zero-offset focusing curves on the acquisition surface, i.e., curves that group traces to focus at different points of the diffractor.

The current paper extends the wave classification and focusing curve building from zero to finite offsets. We illustrate this with a previously unpublished exact solution for the edge diffraction traveltime in a homogeneous medium. In the midpoint-offset representation, we refer to it as the triple-square-root moveout. It complements the set of reference moveouts, which includes the normal moveout of reflections \citep{Levin1971} and the double-square-root moveout of diffractions in 2-D \citep{Landa1987}, also relevant for point diffractions in 3-D. The first part of the paper is dedicated to its derivation, analysis, and verification by wavefield modeling. Next, we derive the general properties of 3-D diffractions and demonstrate them with the triple-square-root moveout. In two appendices, we relate a singularity in the geometrical spreading to the wave classification and derive the NIP-wave theorem necessary to introduce the imaginary eigenwavefronts \citep{Hubral1983} for edge diffractions.

\section{Reference moveouts}
A moveout represents the traveltimes of two-way propagation depending on the source-receiver offset. Reference moveouts are formulae corresponding to different types of scatterers in a homogeneous medium. They have unique characteristics specific to the type of scatterer and provide approximations for realistic models.
Before describing them, let us introduce some basic notions.

In terms of the source and receiver position vectors on the acquisition plane, ${\bf x}_s$ and ${\bf x}_r$, the midpoint, ${\bf x}_m$, and the half-offset vector, ${\bf h}$, are introduced as follows:
\begin{equation}
{\bf x}_m=\frac{{\bf x}_r+{\bf x}_s}{2},~~~~
{\bf h}=\frac{{\bf x}_r-{\bf x}_s}{2}.
\end{equation}
Reversed relations read
\begin{equation}\label{midoff}
{\bf x}_s={\bf x}_m-{\bf h},~~~~
{\bf x}_r={\bf x}_m+{\bf h}.
\end{equation}
With this change of variables, a two-point two-way traveltime $t({\bf x}_s;{\bf x}_r)$ transforms into a function of the midpoint and the half-offset: $\tilde{t}({\bf x}_m;{\bf h})=t({\bf x}_m-{\bf h};{\bf x}_m+{\bf h})$. The tilde symbol over the midpoint-offset representation distinguishes between the two functions. Since they describe the same quantity, we, however, often omit it until there is a need to emphasize the difference.
Given a fixed midpoint, $\tilde{t}({\bf x}_m;{\bf h})$ describes the 3-D moveout.

In 2-D, scalars replace the two-component vectors:
$x_m=(x_r+x_s)/2$ and $h=(x_r-x_s)/2$. To interpret the 3-D moveouts in terms of the 2-D equations, it is helpful to decompose the half-offset vector into a scalar $h$ and an azimuthal vector ${\bf e}_h$: ${\bf h}=h {\bf e}_h$, $\left\lVert{\bf e}_h\right\rVert=1$. 

It is also fruitful to consider local variations of the moveouts due to variations in the midpoint location, i.e., involving midpoints neighboring a central midpoint, ${\bf x}_0$. Particularly, one shifts the midpoint along the same offset direction, such that $\triangle {\bf x}_m={\bf x}_m-{\bf x}_0=m {\bf e}_h$, and studies midpoint-offset traveltime surfaces, $t(m;h)$, that arise for different azimuths.

\subsection{Normal moveout}
The normal moveout (NMO) describes the traveltime for a plane reflector:
\begin{equation}\label{nmo}
t=\sqrt{t_0^2+\frac{4h^2}{v^2_{nmo}}},
\end{equation}
where $t_0$ stands for the zero-offset traveltime and  $v_{nmo}$ denotes a parameter called NMO-velocity. 
It appears, e.g.,
from an application of the cosine law to a triangle between a source, its image, and a receiver.

In a 2-D model with wave velocity $v$ and a straight reflector having dip angle $\alpha$, $v_{nmo}=v/\!\cos\!\alpha$, and the moveout represents a branch of a hyperbola.

\begin{figure}[h!]
\begin{center}
\includegraphics[width=1\columnwidth]{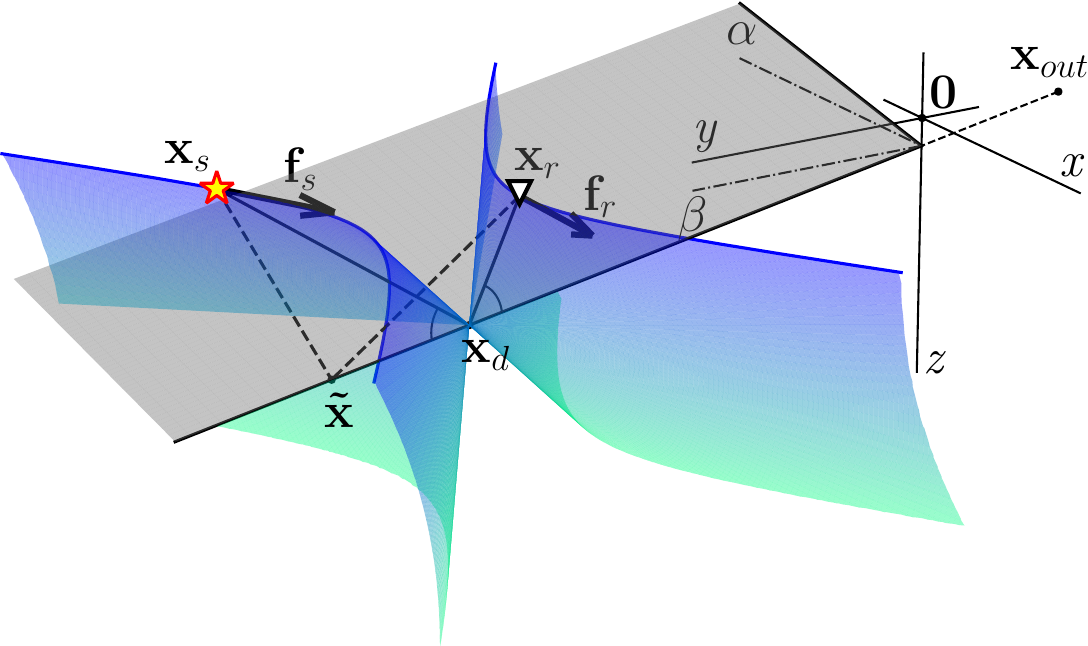}
\end{center}
%\vspace{-0.5cm}
\caption{Diffraction on the edge of a half-plane in a homogeneous isotropic medium. The two-sided diffraction cone for a particular source-receiver pair comprises the incident and diffracted rays. The focusing curves (blue lines) represent the section of the cone by the acquisition plane and can generally be computed using the traveltime measurements.} 
\label{fig:cone}
\end{figure}

In 3-D, the NMO-velocity depends on azimuth: $v^{-2}_{nmo}={{\bf e}_h}^{\!\!T}{\bf S}\,{\bf e}_h$, where ${\bf S}$
is a positive definite matrix. The endpoint of a NMO-velocity vector $v_{nmo}{\bf e}_h$ belongs to an ellipse. Equation \ref{nmo}
describes a sheet of a hyperboloid.
In terms of the apparent dip angles in vertical sections, $\alpha$ and $\beta$ (see Figure \ref{fig:cone}),
\begin{equation}\label{eq2}
{\bf S}=
\frac{1}{v^2}\left(
\setlength\arraycolsep{2pt}
\def\arraystretch{1.2}
\begin{array}{cc}
\displaystyle~~\frac{1+\tan^2\!\beta}{1+\tan^2\!\alpha+\tan^2\!\beta}&\displaystyle-\frac{\tan\!\alpha \tan\!\beta}{1+\tan^2\!\alpha+\tan^2\!\beta}\\
\displaystyle-\frac{\tan\!\alpha \tan\!\beta}{1+\tan^2\!\alpha+\tan^2\!\beta}~&\displaystyle~~\frac{1+\tan^2\!\alpha}{1+\tan^2\!\alpha+\tan^2\!\beta}
\end{array}
\right).
\end{equation}
Aligning the second coordinate axis with the direction of the true reflector dip diagonalizes the matrix:
\begin{equation}
{\bf S}=
\frac{1}{v^2}\left(
\setlength\arraycolsep{2pt}
\begin{array}{cc}
1&0\\
0&~\cos^2\!\beta_0
\end{array}
\right),
\end{equation}
where $\beta_0$ denotes the true dip angle, $\tan^2\!\beta_0=\tan^2\!\alpha+\tan^2\!\beta$.
Expressed through $\beta_0$ and the true dip azimuth $\varphi_0$,  $\tan\!\varphi_0=\tan\!\beta/\!\tan\!\alpha$, the representation \ref{eq2} turns into
\begin{equation}\label{S}
{\bf S}=
\frac{1}{v^2}\left(
\setlength\arraycolsep{2pt}
\def\arraystretch{1}
\begin{array}{cc}
1-\cos^2\!\varphi_0\sin^2\!\beta_0&~-\sin\!\varphi_0\cos\!\varphi_0\sin^2\!\beta_0\\
-\sin\!\varphi_0\cos\!\varphi_0\sin^2\!\beta_0&~1-\sin^2\!\varphi_0\sin^2\!\beta_0
\end{array}
\right).
\end{equation}

Three quantities define the 3-D NMO -- wave velocity, reflector dip, and dipping azimuth. Examining a single moveout surface allows for the estimation of these parameters, except for the dipping azimuth, which is determined up to $180^{\circ}$. This ambiguity occurs due to the possibility of symmetrical reflectors to produce equal moveouts.

It can be removed using the midpoint variation, which alters the zero-offset time in equation \ref{nmo}:
\begin{equation}
t_0({\bf x}_m)=t_0({\bf x}_0)+2{\bf p}^T\!\triangle {\bf x}_m,
\end{equation}
where ${\bf p}$ denotes a column of two horizontal components of the zero-offset upgoing slowness vector.  
For all azimuths, the midpoint-offset surfaces represent elliptic cones with the apexes at zero time, shifted along the $m$-axis to the points of the reflector outcrop. 
In the strike direction or for a horizontal reflector, the apexes go to infinity, and the surfaces unfold into hyperbolic cylinders.

\subsection{Double-square-root moveout}
There is no difference in the kinematics of point and edge diffractions in 2-D. Both follow the double-square-root equation due to the application of the Pythagorean theorem:
\begin{equation}\label{dsr2d}
t=\sqrt{\frac{t_0^2}{4}+\frac{(h^2-2{\hat x}_m h)}{v^2}}+\sqrt{\frac{t_0^2}{4}+\frac{(h^2+2 {\hat x}_m h)}{v^2}},
\end{equation}
where the hat sign over \(x_m\) implies a horizontal coordinate computed relatively to the one of the diffractor, \(\hat{x}_m=x_m-x_d\).

Analogous double-square-root moveout exists for 3-D point diffractions:
\begin{equation}\label{eq4}
t=\sqrt{\frac{t_0^2}{4}+\frac{({\bf h}^2-2 {\hat {\bf  x}}_m^T{\bf h})}{v^2}}+\sqrt{\frac{t_0^2}{4}+\frac{({\bf h}^2+2{ \hat {\bf x}}_m^T{\bf h})}{v^2}},
\end{equation}
where \({ \hat {\bf x}}_m ={\bf x}_m-{\bf x}_d\).

The moveouts \ref{dsr2d} and \ref{eq4} become normal, described by equation \ref{nmo}, with $v_{nmo}=v$ when the midpoint is 
above the diffractor. Furthermore, in 3-D, wherever one shifts the midpoint, the offset direction orthogonal to this shift, ${ \hat {\bf x}}_m^T{\bf e}_h=0$,
equalizes the square roots and makes the moveout normal with $v_{nmo}=v$. Figure \ref{fig:dir1} illustrates such directions and the corresponding NMO-velocities.
Uniqueness of the normalizing azimuth follows
from the non-vanishing of the fourth-order derivative,
\begin{equation}\label{eq5}
\frac{d^4 t^2}{dh^4}(0)\propto({ \hat {\bf x}}_m^T {\bf e}_h)^2\left(\left(\frac{vt_0}{2}\right)^{\!\!2}\!\!-{ \hat {\bf x}}_m^T{\bf e}_h\right),
\end{equation}
for all others. In other directions, the scalar moveout represents equation \ref{dsr2d} with the projection ${\hat {\bf x}}_m^T{\bf e}_h$ playing the role of $\hat{x}_m$.

Similar to the 3-D NMO, the double-square-root moveout \ref{eq4} is determined by three quantities -- wave velocity and two lateral coordinates of the diffractor. One can estimate them from the single moveout surface except for the direction towards the diffractor, which is retrievable up to the $180^{\circ}$ turn. A possibility of symmetrically positioned diffractors to produce equal moveouts causes this ambiguity. 

It disappears if one inserts the variations of the midpoint, ${ \hat {\bf x}}_m=\triangle {\bf x}_m+{ \hat {\bf x}}_0$, and the traveltime,
\begin{equation}
t_0({\bf x}_m)=2\sqrt{\frac{t_0^2({\bf x}_0)}{4}+\frac{(2 { \hat {\bf x}}_0 +\triangle {\bf x}_m)^T\!\triangle {\bf x}_m}{v^2}},
\end{equation}
where \({ \hat {\bf x}}_0 ={\bf x}_0-{\bf x}_d\).
At any azimuth, the midpoint-offset surface represents a square-based ``pyramid'' with smooth and rotationally symmetric apex \citep{Claerbout1985, Alkhalifah2000, Hao2016}. It shifts along the $m$-axis to a point that is an orthogonal projection of the diffractor onto the profile, with $m=-{ \hat {\bf x}}_0^T\!{\bf e}_h$.

\section{Triple-square-root moveout}
\subsection{Derivation and analysis}
The edge diffraction law by \citet{Keller1962} determines the kinematics of waves scattered by a fringe of a finite reflector, a wedge, or a thin elongated body. It is an analogue of Snell's law for reflections, which also follows from Fermat's principle. It states that an incident ray triggers a cone of diffracted rays around the edge with an opening angle equal to the angle between the incident ray and the diffractor (see Figures \ref{fig:cone}, \ref{fig:inhomogeneous_cone}). 

Assume a straight edge in the $y$-$z$ plane of a Cartesian coordinate system (Figure \ref{fig:cone}). Let it have a dip angle $\beta$ and pierce the acquisition plane, $z=0$, at a point ${\bf x}_{out}$ with $x=0$ and $y=y_{out}$, which we call the “outcrop”.

We aim at finding the point of diffraction
migrating along the edge, depending on the source and receiver, similar to the point of reflection that moves along the reflecting boundary. Distance from a point on the acquisition plane with coordinates $x$ and $y$ to the edge reads
\begin{equation}\label{eq6}
r=\sqrt{x^2+(y-y^*)^2+{z^*}^2}=\sqrt{x^2+\sin^2\!\beta(y-y_{out})^2},
\end{equation}
where $y^*$ and $z^*$ denote coordinates of the orthogonal projection of the point onto the edge, $y^*=\cos^2 \! \beta \, y+\sin^2 \! \beta \, y_{out}$, $z^*=\cos \! \beta \sin \! \beta(y-y_{out})$. 
In particular, equation \ref{eq6} describes the distances from source and receiver, $r_s$ and $r_r$. Fermat's principle requires that, subject to a fixed source-receiver pair,  the two-way traveltime to an arbitrary point on the edge $\tilde{{\bf x}}$  with $y=\tilde{y}$,
\begin{equation}\label{eq7}
F=\frac{\sqrt{r_s^2+(\tilde{y}-y_s^*)^2/\!\cos^2 \!\beta}}{v}+\frac{\sqrt{r_r^2+(\tilde{y}-y_r^*)^2/\!\cos^2 \!\beta}}{v},
\end{equation}
gets stationary at the actual point of diffraction, ${\bf x}_d$ with $x=0$ and $y=y_d$: \(\partial F/\partial \tilde{y} (y_d)=0\). This condition 
yields an equation for the diffraction point coordinate,
\begin{equation}\label{eq8}
\frac{y_d-y_s^*}{\sqrt{r_s^2+(y_d-y_s^*)^2/\!\cos^2 \!\beta}}=-\frac{y_d-y_r^*}{\sqrt{r_r^2+(y_d-y_r^*)^2/\!\cos^2 \!\beta}},
\end{equation}
which has a unique solution
\begin{equation}\label{eq9}
y_d=\frac{r_ry_s^*+r_sy_r^*}{r_s+r_r}=\cos^2 \!\beta \,\frac{r_ry_s+r_sy_r}{r_s+r_r}+\sin^2 \! \beta \,y_{out}.
\end{equation}
It represents a sum of
weighted projections or a projection of the sum of weighted coordinates. 

Substitution of equation \ref{eq9} in equation \ref{eq7} and replacing the radii $r_s$ and $r_r$
with expressions like \ref{eq6} lead
to a triple-square-root formula for the diffraction traveltime:
\begin{align}\label{eq10}
~~t=&\frac{1}{v}\sqrt{\cos^2\!\beta \, (y_r-y_s)^2+\left(\sqrt{x_s^2+\sin^2\!\beta(y_s-y_{out})^2}\right.} \nonumber\\
&~~~~~~~~~~~~~~~~~~~~~\overline{\left.+\sqrt{x_r^2+\sin^2\!\beta(y_r-y_{out})^2}\right)^2}.
\end{align}
One can also explain it by using the method of image source. Indeed, the traveltime would not change if one rotates the source around the edge until it comes to the edge-receiver plane on the side opposite the receiver. Then, equation \ref{eq10} follows from the Pythagorean theorem applied to the triangle between the receiver, its image, and the source image.

Due to equation \ref{eq6}, the zero-offset traveltime can be expressed with a quadratic form:
\begin{equation}\label{form}
t_0=2\sqrt{{ \hat {\bf x}}_m^T{\bf \Lambda} { \hat {\bf x}}_m},~~~
{\bf \Lambda}=
\frac{1}{v^2}\left(
\setlength\arraycolsep{2pt}
\begin{array}{cc}
1&0\\
0&~\sin^2\!\beta
\end{array}
\right).
\end{equation}
We use the hat sign over ${\bf x}_m$, similarly to the previous section, to denote a position vector computed relatively to a special point, the point of the edge outcrop in this case: ${ \hat {\bf x}}_m={\bf x}_m-{\bf x}_{out}$. Transforming the inner square roots in equation \ref{eq10}, we obtain the triple-square-root moveout:
\begin{align}\label{tsr}
t=&\sqrt{\frac{4 {h_{\parallel}}^{\!\!\!\!2}}{v^2/\cos^2\!\beta}+\left(\sqrt{\frac{t_0^2}4+({\bf h}-2 { \hat {\bf x}}_m)^T\!{\bf \Lambda}{\bf h}}\right.} \nonumber\\
&~~~~~~~~~~~~~~~~~~~~~~~~\overline{+\left.\sqrt{\frac{t_0^2}{4}+({\bf h}+2{\hat {\bf x}}_m)^T\!{\bf \Lambda}{\bf h}}\right)^{\!\!\!2}},
\end{align}
where $h_{\parallel}$ stands for the edge-parallel offset component. Equation \ref{tsr} has the vector form, i.e., it is valid regardless of the edge orientation. Let ${\bf e}$ denote a vector of the edge azimuth $\varphi$ in the descending direction: ${\bf e}^T=(\cos\!\varphi,\sin\!\varphi)$.
Then, $h_{\parallel}={\bf e}^T{\bf h}$.
Specifically for the setting from Figure \ref{fig:cone},
${\bf e}^T=(0,1)$ and $h_{\parallel}=h_y$.
The positive definite matrix ${\bf \Lambda}$ accordingly transforms: 
\begin{equation}\label{general_lambda}
{\bf \Lambda}=
\frac{1}{v^2}\left(
\setlength\arraycolsep{2pt}
\begin{array}{cc}
1-\cos^2\!\varphi\cos^2\!\beta&~-\sin\!\varphi\cos\!\varphi\cos^2\!\beta\\
-\sin\!\varphi\cos\!\varphi\cos^2\!\beta&~1-\sin^2\!\varphi\cos^2\!\beta
\end{array}
\right).
\end{equation}
It differs from the matrix ${\bf S}$ (equation \ref{S}) by the presence of the cosine of the dipping angle instead of the sine. It is noteworthy that $\bf \Lambda$ implements the following decomposition:
\begin{equation}
\frac{h^2}{v^2}=\frac{{h_{\parallel}}^{\!\!\!\!2}}{v^2/\cos^2\!\beta}+{\bf h}^{\!T}\!{\bf \Lambda}{\bf h}.
\end{equation}

The triple-square-root moveout \ref{tsr} can be viewed as the 2-D NMO (equation \ref{nmo}) relative to $h_{\parallel}$ with the dip angle $\beta$ and the generalized double-square-root replacing $t_0$. 
In this precise sense, the edge diffraction is a composition of diffraction and reflection. Moreover, for a horizontal edge, 
the generalized double-square-root reduces to the double-square-root \ref{dsr2d} relative to the edge-orthogonal offset component $h_{\perp}$:
\begin{align}\label{eq13}
t=&\sqrt{\frac{4 {h_{\parallel}}^{\!\!\!\!2}}{v^2}+\left(\sqrt{\frac{t_0^2}{4}+\frac{(h_{\perp}^2-2 \hat{x}_m h_{\perp})}{v^2}}\right.}\nonumber\\
&~~~~~~~~~~~~~~~~~~~~~~~~\overline{+\left.\sqrt{\frac{t_0^2}{4}+\frac{(h_{\perp}^2+2{\hat x}_m h_{\perp})}{v^2}}\right)^2}, 
\end{align}
where $\hat x_m=x_m-x_{out}$.

Turning to the description of special azimuths, we first consider an acquisition line orthogonal to the edge. In the initial coordinate system, it means that $h_{\parallel}=h_y=0$ and
$y_d=y_s^*=y_r^*=const.$ The moveout reduces to the double-square-root \ref{dsr2d} with $\hat x_m=x_m-x_{out}$. Uniqueness of the direction causing the double-square-root moveout can be proved by equating
formula \ref{tsr}, with substituted ${\bf h}=h{\bf e}_h$, to formula \ref{dsr2d}, with an unknown parameter instead of $\hat x_m$. Squaring this identity three times and setting coefficients of the resulting fourth-order polynomial in $h$ to zero results in ${\bf e}^T\!{\bf e}_h=0$.

\begin{figure*}
\center
%\hfill
\subfigure[]{\includegraphics[width=0.26\linewidth]{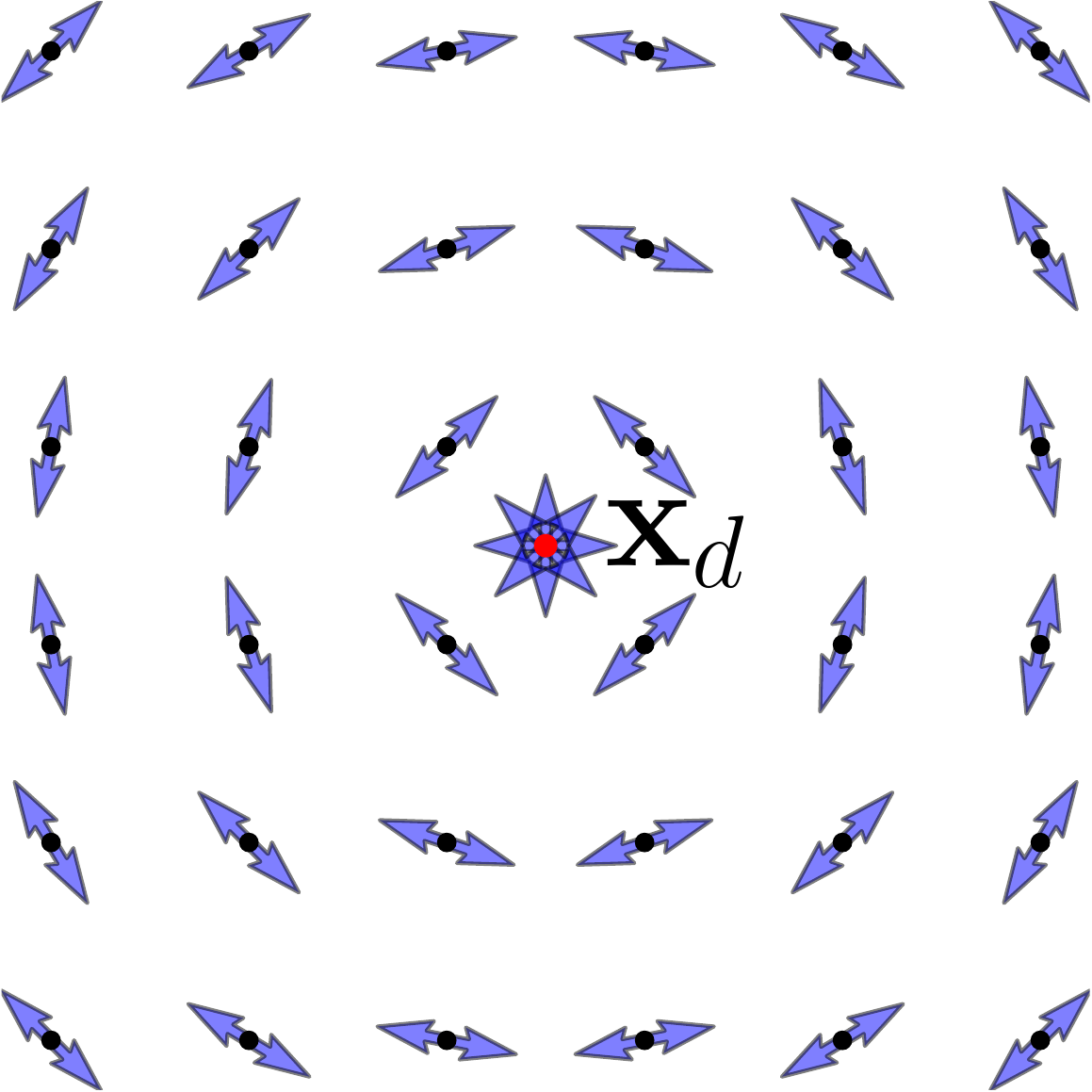}\label{fig:dir1}}
\hspace{0.05\linewidth}
\subfigure[]{\includegraphics[width=0.26\linewidth]{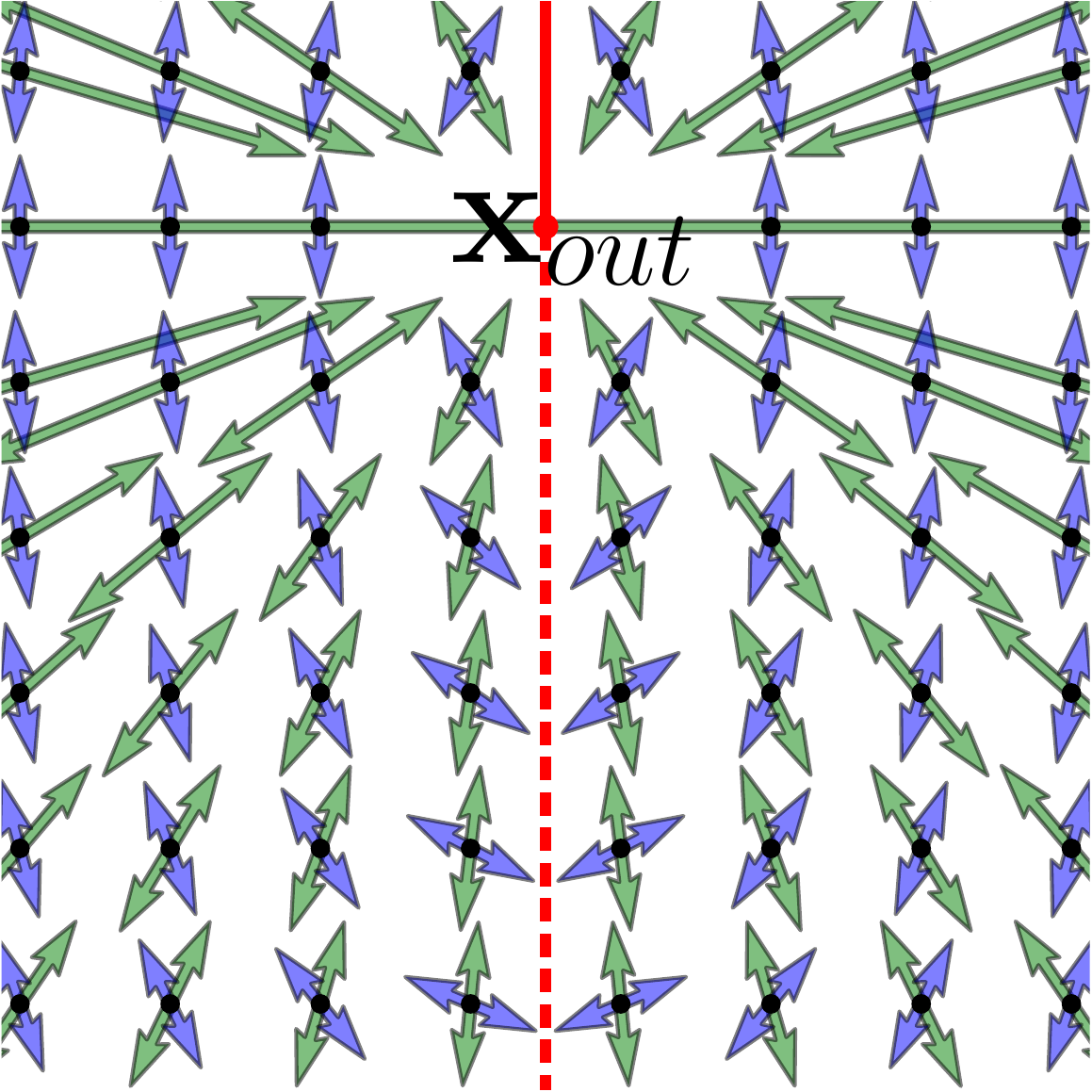}\label{fig:dir2}}
\hspace{0.05\linewidth}
\subfigure[]{\includegraphics[width=0.26\linewidth]{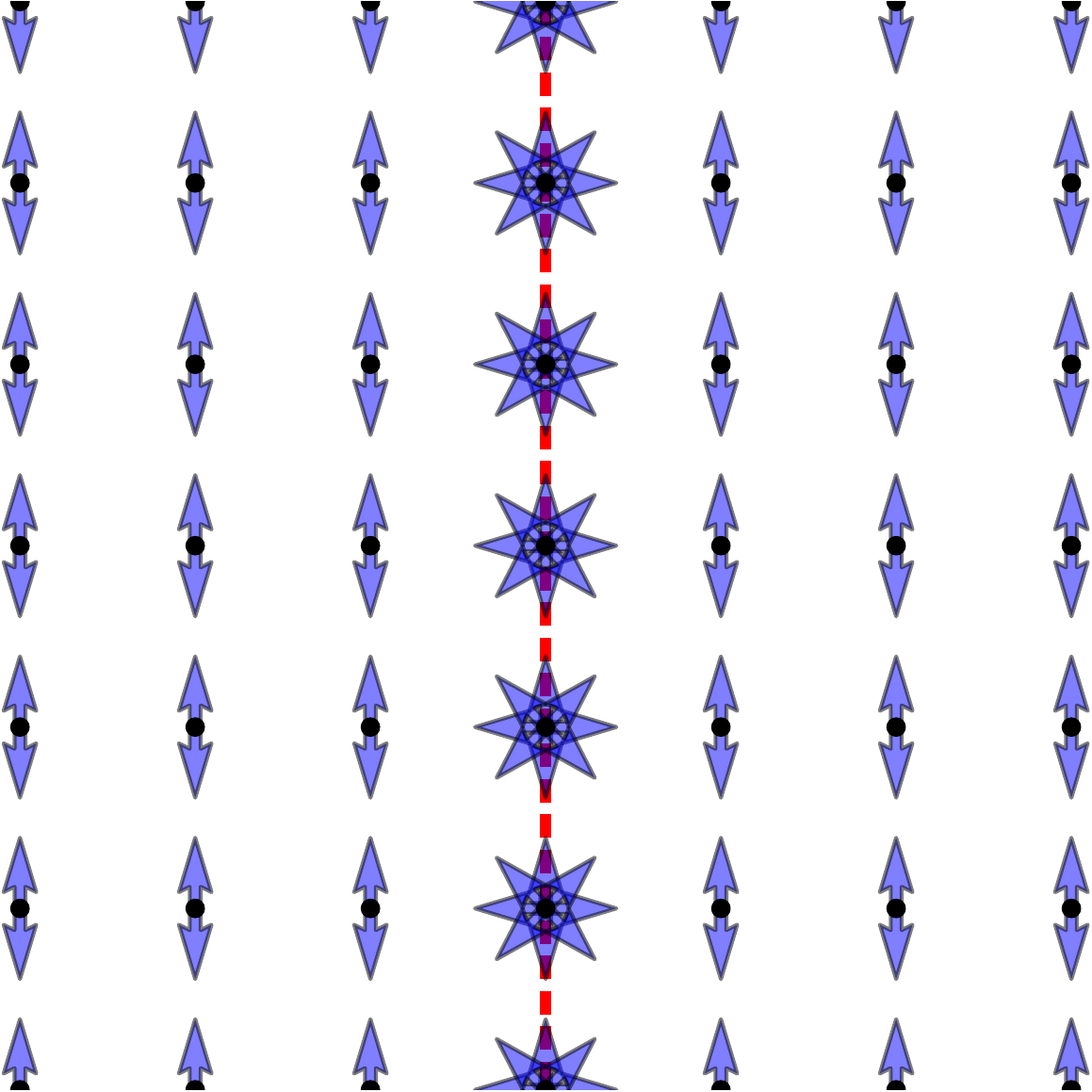}\label{fig:dir3}}
\caption{Top view on azimuths and magnitudes ($v_{nmo}$) of normal moveouts for: point diffractor (\subref{fig:dir1}); inclined edge diffractor with dip angle $\beta=30^{\circ}$ (\subref{fig:dir2}); horizontal edge diffractor (\subref{fig:dir3}). }
\label{fig:directions}
\end{figure*}

For edge diffractions, there also exist normalizing azimuths, i.e., directions that cause hyperbolic moveout. In contrast to point diffraction (Figure \ref{fig:dir1}), there are two such families, illustrated in Figure \ref{fig:dir2}. 

Similar to point diffraction, aligning the offset in directions of the first kind (blue arrows) equalizes the inner square roots in equation \ref{tsr} and leads to $v_{nmo}=v$. A condition for the radicals to become equal reads
\begin{equation}\label{maintext_condition}
{ \hat {\bf x}}^T_m {\bf \Lambda} {\bf e}_h=0.
\end{equation} 
In the coordinate system from Figure \ref{fig:cone}, it simplifies to
an equation
\begin{equation}
x_m ({\bf e}_h)_x+\sin^2\!\beta  (y_m-y_{out}) ({\bf e}_h)_y=0,
\end{equation}
which can be solved with respect to ${\bf e}_h$.  The surface components of the zero-offset slowness vector follow from differentiating the zero-offset traveltime \ref{form}: ${\bf p}=t_0^{-1}\!{\bf \Lambda}{ \hat {\bf x}}_m$. Thus, condition \ref{maintext_condition} implies orthogonality of the azimuthal vector to the zero-offset ray: ${\bf p}^T\!{\bf e}_h=0$. Furthermore, since the radicals are equal, $r_s=r_r$, $y_d=y_m^*=const.$, according to equation \ref{eq9}. Increasing the offset along such a direction does not alter the diffraction point.

The directions of the second kind (green arrows in Figure \ref{fig:dir2}) point towards the outcrop: ${\bf h}\!\parallel\!{\hat {\bf x}}_m$. To explain them, consider an arbitrary plane containing the edge. 
It intersects the acquisition plane by a straight line passing through the outcrop. For a midpoint on this line, aligning the offsets in this direction causes diffracted rays exclusively in the aforementioned skew plane. As a result, the kinematics becomes
equivalent to the ones of 2-D reflection. 
In this case, complete squares appear under each of the inner roots in equation \ref{tsr}. 
Away from the outcrop, when $\left\lVert { \hat {\bf x}}_m\right\lVert>\left\lVert {\bf h} \right\lVert$,
\begin{equation}\label{skew_nmo}
v_{nmo}^2=\frac{v^2}{\cos^2\!\beta \cos^2\!\theta}
\end{equation}
with $\theta$ being an angle between ${ \hat {\bf x}}_m$ and ${\bf e}$. $\cos\!\beta \cos\!\theta$ represents the cosine of the edge dip angle in the skew plane defined by the edge and the profile.  

To verify that no other direction causes the hyperbolic moveout, we again refer to the fourth-order derivative:
\begin{equation}\label{eq15}
\frac{d^4 t^2}{dh^4}(0)\propto \sin^2\!\beta ({ \hat {\bf x}}^T_m {\bf \Lambda}{\bf e}_h)^2({ \hat {\bf x}}_m\!\!\times\!{\bf e}_h)^2.
\end{equation}

The case of a horizontal diffractor stands out (see Figure \ref{fig:dir3}). Outside the edge projection (dashed red line), the two families merge, becoming parallel to the edge. Precisely above the diffractor, any azimuth causes the hyperbolic moveout (see equation \ref{eq13}). $v_{nmo}=v$ everywhere, same as for a point diffractor. 

In contrast to the reference 3-D moveouts from the previous section, the triple-square-root \ref{tsr} requires four independent parameters to be specified -- wave velocity $v$, edge dip angle $\beta$, edge azimuth in the dipping direction $\varphi$, and azimuth of the outcrop-midpoint direction relative to the edge $\theta$. Examining the moveout surface at a single midpoint allows one to determine all four parameters, but the inclination azimuth and the direction to the outcrop only up to $180^\circ$. The following procedure is feasible.
First, determine the two normalizing azimuths, along with the corresponding NMO-velocities. A direction with a smaller velocity is one of the first kind. Let us denote the corresponding unit vector ${\bf e}^n_{1}$. It brings the value of $v$.  The remaining  direction of the second kind with larger velocity, ${\bf e}^n_{
2}$, passes through the outcrop. However, we cannot determine on which side. Equation \ref{skew_nmo} and the value of $v$
yield a product $\cos^2\!\beta\cos^2\!\theta$. Aligning the Cartesian system with ${\bf e}^n_2$, such that  $({\bf e}_2^n)_2=0$, allows to replace $\varphi$ with $\theta$ in the matrix \ref{general_lambda} and rewrite
equation \ref{maintext_condition}:
\begin{equation}
{{\bf e}_2^n}^T\!{\bf \Lambda}{\bf e}^n_1=(1-\cos^2\!\theta \!\cos^2\!\!\beta)({\bf e}^n_1)_1-\sin\!\theta\!\cos\!\theta\!\cos^2\!\!\beta({\bf e}^n_1)_2=0.
\end{equation}
Dividing it by $\cos^2\!\theta\cos^2\!\beta$, we express $\tan\!\theta$ in terms of the known quantities:
\begin{equation}
\tan\!\theta=\left(\frac{1}{\cos^2\!\beta\cos^2\!\theta}-1\right)\!\frac{({\bf e}^n_1)_1}{({\bf e}^n_1)_2}.
\end{equation}
This way, $\theta$ is determined up to $180^{\circ}$, which also yields  $\varphi$ and $\cos\!\beta$. $\varphi=\varphi_0+\theta$, where $\varphi_0$ is the azimuth of ${\bf e}^n_2$ before the coordinate rotation.
There are four options: two opposite directions toward the outcrop and two opposite possibilities for the inclination direction. In practice, we are dealing with buried diffractors, whose hypothetical outcrops are far away. Hence, it is reasonable to assume ${\bf e}^T{ \hat {\bf x}}_m>0$, which leaves two symmetrical positions: the outcrop can be on one side or the other with the edge inclining to the midpoint direction.

One of them can be further discarded by involving the neighboring midpoints, which means a substitution of ${ \hat {\bf x}}_m=\triangle{\bf x}_m+{\hat {\bf x}}_0$ and 
\begin{equation}
t_0({\bf x}_m)=\sqrt{t_0^2({\bf x}_0)+4(2{ \hat {\bf x}}_0 +\triangle {\bf x}_m)^T\! {\bf \Lambda} \triangle {\bf x}_m}
\end{equation}
into equation \ref{tsr}. In comparison to reflection and point diffraction,
there are more options for the shape of the midpoint-offset surface.
In the direction orthogonal to the edge, it represents the “pyramid”, typical of point diffraction. According to \citet{Znak2023}, for a general edge diffraction, which includes heterogeneous anisotropic media and curvilinear edges, there is an azimuth where the curvatures of the midpoint-offset surface become equal.
Along the edge, the midpoint-offset surface is given by the triple-square-root, which simplifies to the elliptic cone of reflection above the edge and the hyperbolic cylinder for a horizontal edge. The elliptic cones also appear along the normalizing directions of the second kind due to the reflection-like kinematics in the skew planes. Apexes of these cones shift by the distance from the central midpoint to the outcrop. However, along the normalizing directions of the first kind, the midpoint-offset surfaces represent sheets of hyperboloids:
\begin{equation}
t=\sqrt{t_0^2+4{{\bf e}^n_1}^T\!\!{\bf \Lambda} {\bf e}^n_1 m
^2+\frac{4h^2}{v^2}},
\end{equation}
\begin{equation}
{{\bf e}^n_1}^T\!\!{\bf \Lambda} {\bf e}^n_1=\frac{1}{v^2}\frac{\sin^2\!\beta \sin^2\!\theta+\sin^4\!\beta \cos^2\!\theta}{\sin^2\!\theta+\sin^4\!\beta \cos^2\!\theta}\leqslant{\frac{1}{v^2}}.
\end{equation}

\subsection{Wavefield simulation}
\begin{figure*}
\center
%\hfill
\subfigure[]
{\includegraphics[width=0.288\linewidth]{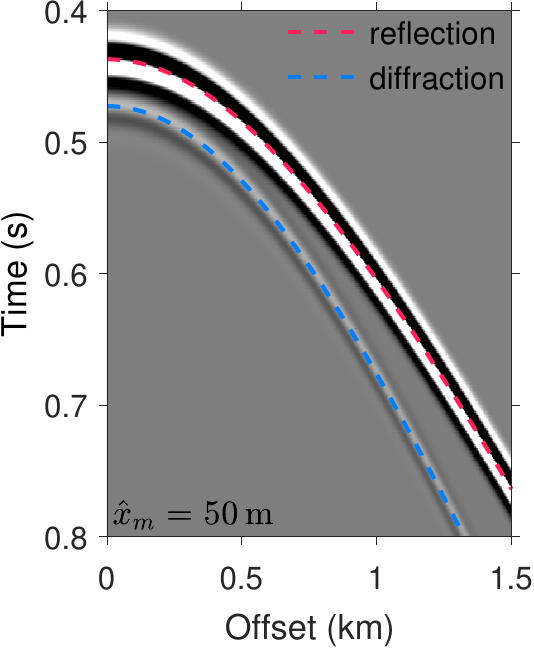}\label{fig:wave1}}
%\hfill
\hspace{0.025\linewidth}
\subfigure[]
{\raisebox{+0.01cm}{\includegraphics[width=0.28\linewidth]{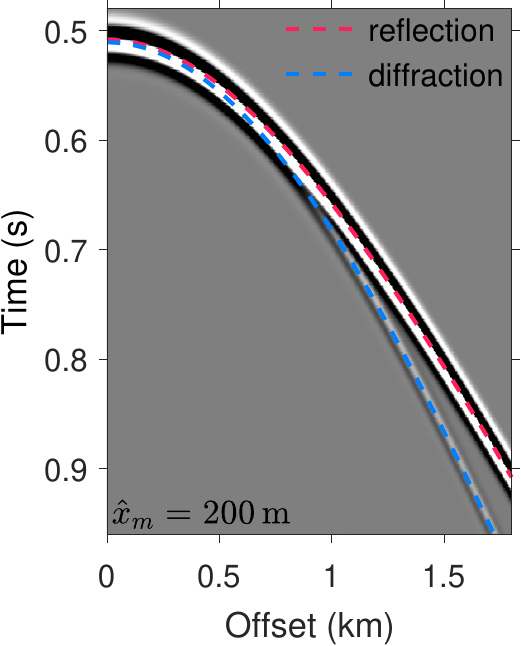}\label{fig:wave2}}}
%\hfill
\hspace{0.021\linewidth}
\subfigure[]
{
\raisebox{+0.05cm}{
\includegraphics[width=0.278\linewidth]{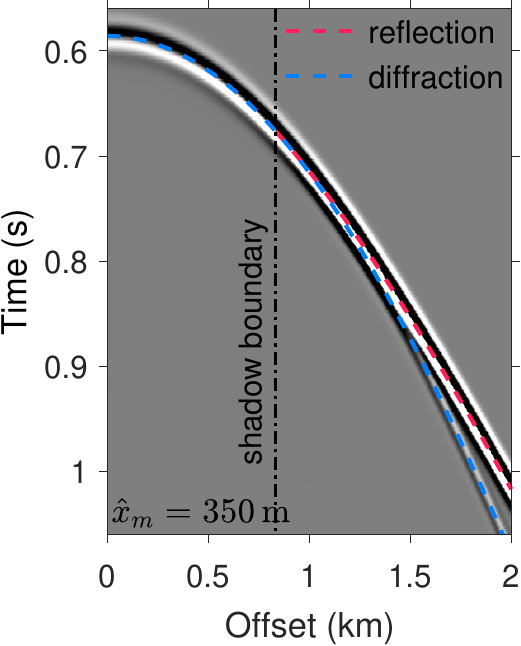}\label{fig:wave3}}}
%\hfill
\caption{Common-midpoint seismograms for the model in Figure \ref{fig:cone}. The midpoints lay on a line orthogonal to the edge: roughly above the edge (\subref{fig:wave1}); near the zero-offset shadow boundary (\subref{fig:wave2}); further away (\subref{fig:wave3}). The offsets are oriented at an azimuth of $45^{\circ}$.}
\label{fig:waves}
\end{figure*}

In this section, we verify the triple-square-root moveout by wave modeling. We simulate common-midpoint gathers (Figure \ref{fig:waves}) in a 3-D homogeneous medium comprising a contrast half-plane, as shown in Figure \ref{fig:cone}. The scattering layer 
of $10$ m thickness consists of a material with $2500$ m/s wave velocity. The wave velocity in the surrounding matter is $2000$ m/s.
The half-plane apparent dip angles, $\alpha$ and $\beta$, equal to $30^{\circ}$ and $20^{\circ}$. As the source time function, we choose the Ricker wavelet with a central frequency of $30$ Hz. The waveforms are computed based on the Born approximation for the wave equation 
\citep[see, e.g., equation 9.3.7 in][ for its zero-offset version in the frequency domain]{Bleistein1984}. 

We consider a line on the acquisition surface orthogonal to the edge, along the $x$-axis, such that the edge is $500$ m below. On this line, we select three representative midpoints at different distances from the reflector: $\hat{x}_m=50$ m, $200$ m, and $350$ m. The offset azimuths are $45^{\circ}$. By this choice of geometry, we intend to demonstrate different mutual arrangements of the reflected and diffracted signals.

Figure \ref{fig:waves} demonstrates the modeled waveforms overlayed with the traveltimes of reflection (red dashed line) and edge diffraction (blue dashed line), which we compute using formulas \ref{nmo} and \ref{tsr}. Due to the shift of the midpoint, the seismograms are in different regimes. Close to the edge, at $x_m=50$ m, the signals of reflection and diffraction do not overlap (Figure \ref{fig:wave1}). When the midpoint moves closer to the zero-offset shadow boundary, to $x_m=200$ m, the signals partially overlap around the zero offset and diverge at longer offsets (Figure \ref{fig:wave2}). 
When the midpoint moves deeper into the zero-offset shadow, to $x_m=350$ m, the reflection disappears at shorter offsets, and the edge diffraction prevails (Figure \ref{fig:wave3}). Note the phase reversal of the edge diffraction that takes place at the shadow boundary \citep{Klem2008}. The order of black and white phases changes to white and black. Generally, the triple-square-root moveout efficiently predicts the edge diffraction traveltimes.

\section{Wave type signature} 
3-D reflections, point diffractions, and edge diffractions have unique signatures encoded in their traveltimes, which allow a data-driven wave type identification. This approach to characterization is local, in the sense that we analyze the second-order traveltime derivatives for a chosen source-receiver pair. However, it requires the 4-D source-receiver space to be involved. In other words, it is based on the local moveout variation.

Consider a two-point two-way traveltime on the acquisition plane $t({\bf x}^{(1)}; {\bf x}^{(2)})$ and a matrix ${\bf D}$ of its mixed source-receiver derivatives at ${\bf x}^{(1)}={\bf x}_s$ and ${\bf x}^{(2)}={\bf x}_r$:
\begin{equation}
D_{ij}=2\frac{\partial^2t}{\partial x_i^{\!(1)}\! \partial x_j^{\!(2)}}({\bf x}_s;{\bf x}_r),~~~~~~i,j=1,2.  \label{eq1_a}
\end{equation}
It is non-symmetric, and swapping the source and receiver results in its transposition. We call ${\bf D}$ the ``diffraction identification matrix'' since its algebraic properties indicate the wave type. It degenerates for diffractions of both types. Moreover, the degree to which it is degenerate determines the type of diffraction. The multiplier in equation \ref{eq1_a} is optional. We need it to reconcile with a definition previously introduced by \citet{Znak2023} for the case of no offset.

We first address point diffractions. Denote the two-point direct traveltime as $\tau({\bf x};{\bf y})$. Its first vector argument, ${\bf x}$, 
with two components, describes a point on the acquisition plane. The second argument, ${\bf y}$, is a three-component position vector in the subsurface. Then, the two-point two-way traveltime for a diffractor at ${\bf y}={\bf x}_d$ reads
\begin{equation}
t({\bf x}_s; {\bf x}_r)=\tau({\bf x}_s;{\bf x}_d)+\tau({\bf x}_r;{\bf x}_d).
\end{equation}
The slowness vector components at one branch are independent of the position of the second branch:
\begin{equation}
\frac{\partial t}{\partial x_j^{\!(2)}}({\bf x}_s;{\bf x}_r)=\frac{\partial \tau~}{\partial x_j}({\bf x}_r; {\bf x}_d),~~~~~~j=1,2.
\end{equation}
\citet{Bauer2016} emphasize the importance of this type of source-receiver decoupling in diffraction processing.
Applying an additional differentiation with respect to the source results in
\begin{equation}
\frac{\partial^2t}{\partial x_i^{\!(1)} \!\partial x_j^{\!(2)}}({\bf x}_s;{\bf x}_r)=0,~~~~~~i,j=1,2.
\end{equation}
We conclude that the identification matrix vanishes, ${\bf D}={\bf 0}$. In this way, it fully degenerates.

\begin{figure}[h]
\center{\includegraphics[width=1\linewidth]{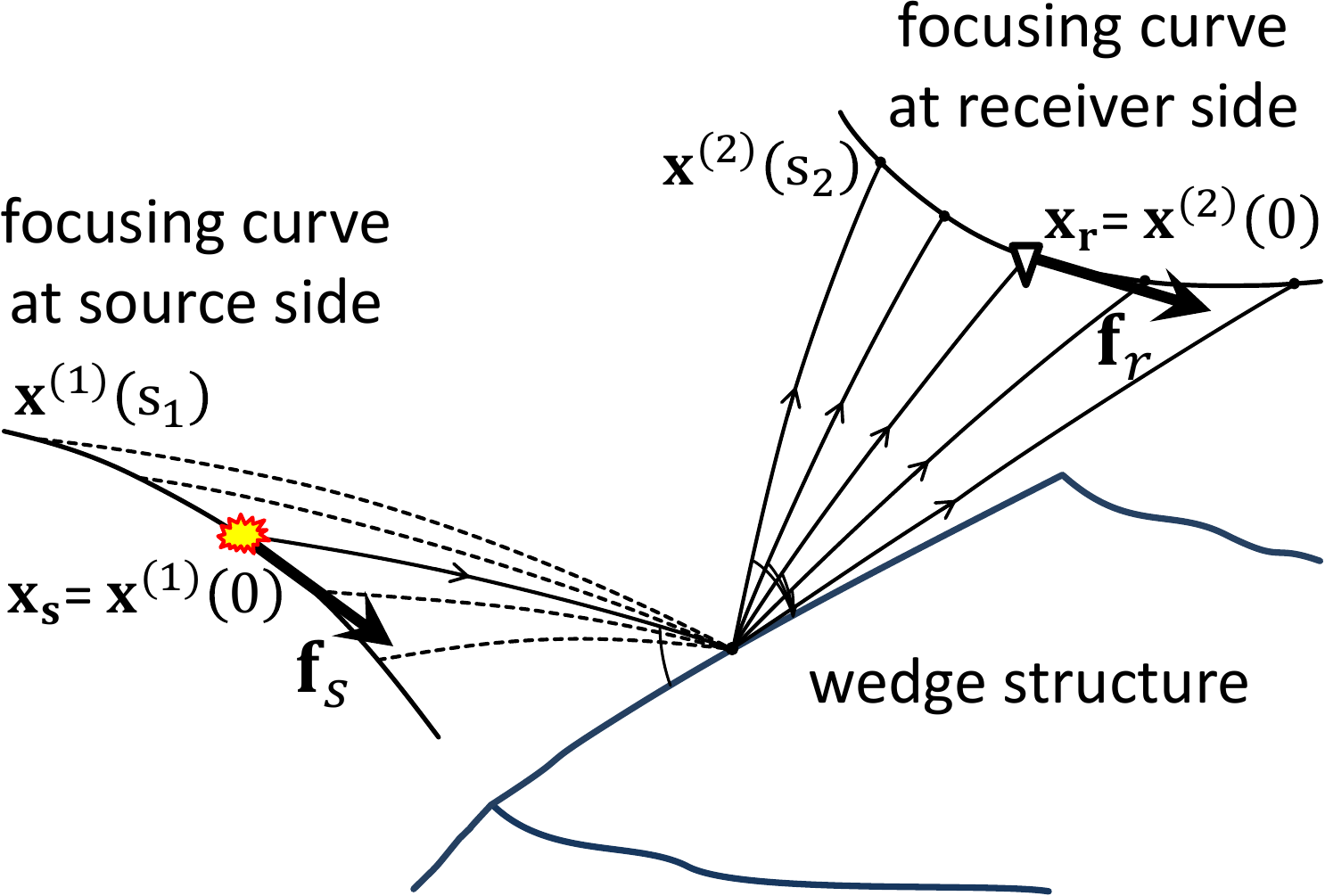}} 
\caption{The double diffraction cone for a specific source-receiver pair in a heterogeneous isotropic medium. At the receiver side, it is formed by the actual diffracted rays (solid lines). At the source side, it consists of potentially diffracted rays as if the source and receiver are interchanged (dashed lines).}\label{fig:inhomogeneous_cone}
\end{figure}

Edge diffractions are characterized according to \citet{Keller1962}. An incident ray generates diffracted rays that form a cone around the edge, as illustrated in Figures \ref{fig:cone} and \ref{fig:inhomogeneous_cone}. The opening angle of this cone equals the angle between the incident ray and the edge. This description allows for curvilinear diffractors. In heterogeneous media, Keller's law is applicable locally at the point of incidence. The ray bending caused by the heterogeneity deforms the conical surfaces. A more general formulation implies conservation of the edge-tangent component of the slowness vector. This principle holds for anisotropic media \citep{Tygel1999}, where the diffraction cones are generally not circular and asymmetric even under homogeneous conditions.

The one-sided cone postulated by Keller’s law is insufficient for our analysis. We require an extended notion of double cone (see Figures \ref{fig:cone}, \ref{fig:inhomogeneous_cone}), which naturally arises when one thinks of multiple sources. The rays of the diffraction cone at the receiver side share the same angle to the edge. Hence, each of them could potentially excite another cone of rays on the source side when reversed. These two parts combine to form the double cone for a given source-receiver pair. The double cone intersects the acquisition plane by two lines that we call “focusing curves”. One focusing curve passes through the receiver and another -- through the source (Figures \ref{fig:cone}, \ref{fig:inhomogeneous_cone}). Rays propagating from them into the subsurface, with corresponding time intervals and initial slowness vectors,
focus at a single point on the edge, the apex of the cone. In the zero-offset case, the double cone collapses into a deformed plane orthogonal to the edge,  a single cone in the case of anisotropy, and the two focusing curves merge \citep{Znak2023}.  

Let us denote the focusing curves as ${\bf x}^{(1)}(s_1)$ and ${\bf x}^{(2)}(s_2)$ and assume their parameterization with arc lengths $s_1$ and $s_2$, such that ${\bf x}^{(1)}(0)={\bf x}_s$ and ${\bf x}^{(2)}(0)={\bf x}_r$ (Figure \ref{fig:inhomogeneous_cone}).
Substituting these into the two-point two-way edge diffraction time, $t({\bf x}^{(1)}(s_1);{\bf x}^{(2)}(s_2))$, limits propagation to the double cone only. In this way, we can also restrict the slowness vector components. Calculated on one side of the double cone, they do not depend on the position of the branch on the other side: 
\begin{equation} \frac{\partial t}{\partial x_j ^{\!(2)}}({\bf x}^{(1)}(s_1);{\bf x}^{(2)}(s_2))= \frac{\partial \tau}{\partial x_j}({\bf x}^{(2)}(s_2);{\bf x}_d),~~~~~~j=1,2, \label{eq5_a} 
\end{equation}
where ${\bf x}_d$ denotes the diffraction point on the edge for the particular ${\bf x}_s$ and ${\bf x}_r$.
Indeed, moving the source along the source-side focusing curve preserves the angle between the incident ray and the edge. As a result, the configuration of rays triggered at the receiver side remains unchanged. In general, there is no decoupling of the source and receiver, as occurs for point diffractions. However, reducing the source-receiver space to the focusing curves fixes the diffraction point. Within these curves, the decoupling does occur.
Differentiating equation \ref{eq5_a} with respect to $s_1$ and setting $s_1$ and $s_2$ to zero yields
\begin{equation}
\frac{\partial^2t}{\partial x_i^{\!(1)} \!\partial x_j^{\!(2)}}({\bf x}_s;{\bf x}_r)\frac{d x_i}{ds_1}^{\!\!(1)}\!\!\!\!\!(0)=0,~~~~~~j=1,2,
\label{eq6_a}
\end{equation}
which is written in vector form as follows
\begin{equation}\label{mateqs}
{\bf D}^T\!{\bf f}_s={\bf 0}.
\end{equation}
Here, we denote the unit vector tangent to the focusing curve at the source side as ${\bf f}_s$. 
The same reasoning applies in reverse.  Specifically, by considering the slowness vector at the source side, which is independent of the position of the branch on the receiver side, we obtain
\begin{equation}
\frac{\partial^2t}{\partial x_j^{\!(1)} \!\partial x_i^{\!(2)}}({\bf x}_s;{\bf x}_r)\frac{d x_i}{ds_2}^{\!\!(2)}\!\!\!\!\!(0)=0,~~~~~~j=1,2
\label{eq7_a}
\end{equation}
or
\begin{equation}\label{mateqr}
{\bf D}{\bf f}_r={\bf 0},
\end{equation}
where the unit tangent to the focusing curve at the receiver side is denoted as ${\bf f}_r$. Equations \ref{mateqs}, \ref{mateqr} are consistent with the equivalence of the identification matrix transposition and the source-receiver swapping.

Equations \ref{eq6_a}-\ref{mateqr} establish linear dependencies among the rows and columns of the identification matrix. This reduces the rank of the matrix ${\bf D}$ to one, making it singular -- so $\mbox{det}{\bf D}=0$ while ${\bf D}\ne{\bf 0}$. For point diffractions, $\mbox{det}{\bf D}=0$ also holds. We see that this property is shared by both types of diffraction. Therefore, by analyzing the determinant, one can distinguish reflections (where $\mbox{det}{\bf D}\ne0$) from diffractions (where $\mbox{det}{\bf D}=0$). Furthermore, examining the matrix elements enables differentiation between points and edges. The degeneracy of the identification matrix closely relates to a singularity in the two-way geometrical spreading (see Appendix \ref{ap:0}).

To switch from the source-receiver to the midpoint-offset description, it is necessary to 
relate the second-order derivatives. Differentiating the identity $t({\bf x}_s;{\bf x}_r)=\tilde{t}({\bf x}_m({\bf x}_s,{\bf x}_r);{\bf h}({\bf x}_s,{\bf x}_r))$ leads to the following representation for the identification matrix:
\begin{align}\label{Dxh}
D_{ij}
=\frac{1}{2}&\left(\frac{\partial^2 t}{\partial x_i^m \partial x_j^m}
-\frac{\partial^2 t}{\partial h_i \partial h_j}+\frac{\partial^2 t}{\partial x_i^m \partial h_j}-\frac{\partial^2 t}{\partial x_j^m \partial h_i}\right),\nonumber\\
&~~~~~~~~~~~~~~~~~~~~~~~~~~~~~~~~~~~~~~~~~i,j=1,2.
\end{align}
It is decomposed into a symmetric part, consisting of the first two terms, and an antisymmetric part, comprising the last two terms. 

The left-hand side of equation \ref{Dxh} must represent the zero matrix for point diffractions. Because it is symmetric and the decomposition is unique, both the symmetric and antisymmetric parts disappear. Note that the presence of the second condition is exclusive to 3-D. In 2-D, the antisymmetric part generally does not exist.

\begin{figure}[h]
\center{\includegraphics[width=1\linewidth]{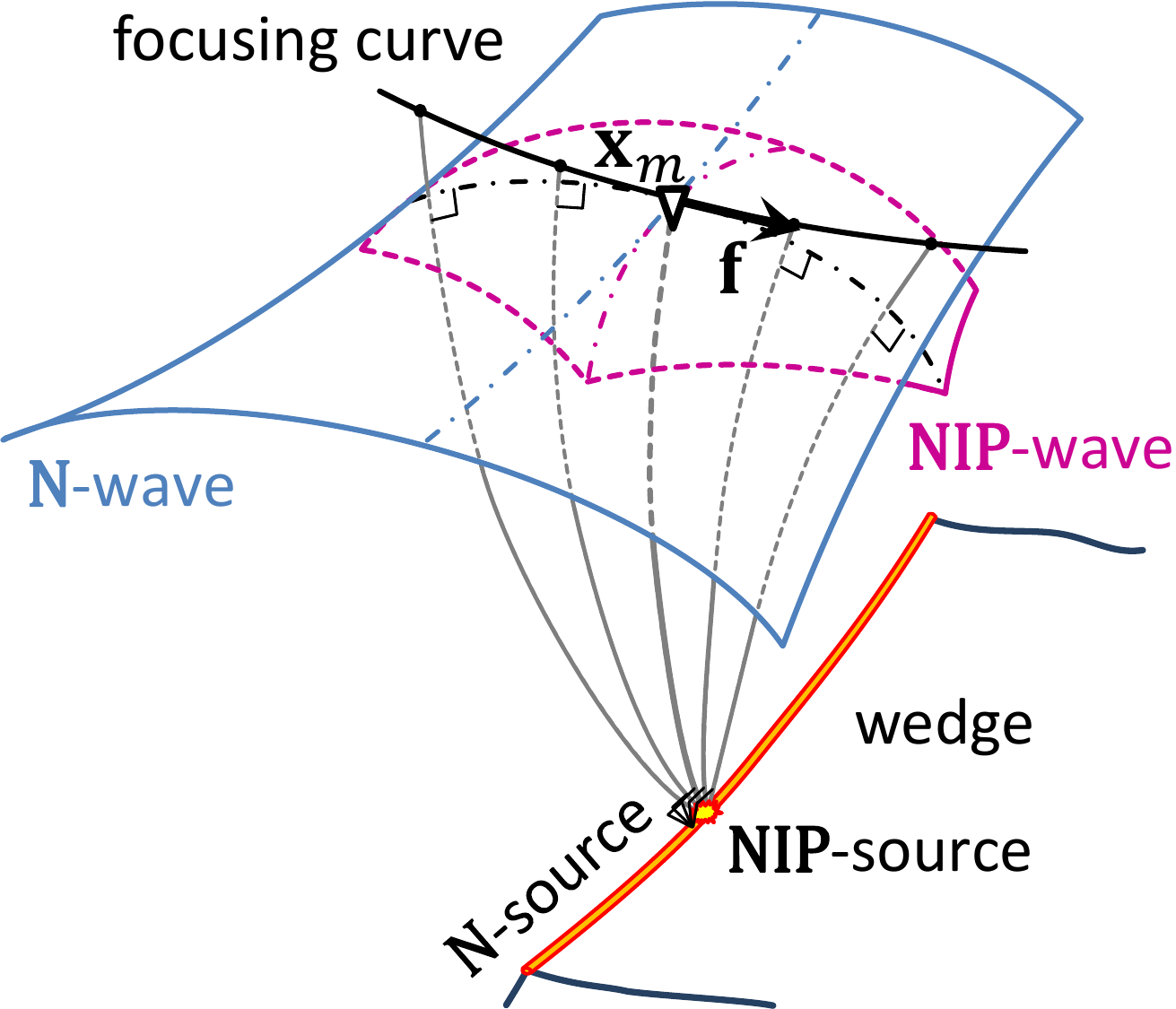}} 
\caption{The imaginary eigenwavefronts in the case of an edge diffractor in an isotropic medium: N-wave caused by excitation of the edge line, NIP-wave due to a source at the point of normal incidence of the zero-offset ray.}\label{fig:eigenwaves}
\end{figure}

Due to the traveltime reciprocity, it also vanishes at zero offset regardless of the wave type:
\begin{equation}\label{matrix_zo}
D_{ij}
=\frac{1}{2}\left(\frac{\partial^2 t}{\partial x_i^m \partial x_j^m}
-\frac{\partial^2 t}{\partial h_i \partial h_j}\right),~~~~~~i,j=1,2.
\end{equation}
The remaining terms can be expressed through a pair of imaginary upgoing wavefronts (see Figure \ref{fig:eigenwaves}). These are eigenwavefronts introduced by \citet{Hubral1983} for reflections. The first N-wave, with traveltime $t^N$, could be hypothetically caused by simultaneous excitation of the scatterer.  In our case, this is an edge line. The second imaginary NIP-wave, with traveltime $t^{NIP}$,  could be triggered by a point source at the point of zero-offset normal incidence (NIP). A substitution for the first term, 
\begin{equation}\label{substitution}
\frac{\partial^2 t}{\partial x_i^m \partial x_j^m}({\bf x}_m;{\bf 0})=2\frac{\partial^2 t^N}{\partial x_i \partial x_j}({\bf x}_m),~~~~~~i,j=1,2,
\end{equation}
follows from a relation $t^N({\bf x}_m)=t({\bf x}_m;{\bf 0})/2$. A substitution for the second term is based on the NIP-wave theorem. First formulated for reflections \citep{Krey1976, Chernjak1979, Hubral1983}, it affirms that
\begin{equation}\label{theorem_maintext}
\frac{\partial^2 t}{\partial h_i \partial h_j}({\bf x}_m;{\bf 0})=2 {\frac{\partial^2 t}{\partial x_i \partial x_j}}^{\!\!\!\!\!\!\!NIP}\!\!\!\!\!({\bf x}_m;{\bf x}_m)~~~i,j=1,2.
\end{equation}
Here, we display two arguments of the NIP-wave traveltime: the time field variable on the acquisition plane and the midpoint that is necessary to define a NIP-wave. The derivatives on the right-hand side of equation \ref{theorem_maintext} are calculated with respect to the first argument. In Appendix \ref{ap:1}, we derive equation \ref{theorem_maintext} for a general curvilinear diffractor in a heterogeneous anisotropic medium. However, with the exact formula \ref{tsr}, one can verify it directly. This requires the traveltime of the spherical NIP-waves:
\begin{equation}
t^{NIP}(x,y;x_m,y_m)=\frac{\sqrt{x^2+(y-y_m^*)^2+{z_m^*}^{\!\!2}}}{v}.
\end{equation}
As a result of the substitutions \ref{substitution} and \ref{theorem_maintext}, the zero-offset identification matrix reduces to a difference between the second-order traveltime derivatives of the eigenwavefronts:
\begin{equation}\label{eq18}
D_{ij}={\frac{\partial^2 t}{\partial x_i \partial x_j}}^{\!\!\!\!\!\!\!N}-{\frac{\partial^2 t}{\partial x_i \partial x_j}}^{\!\!\!\!\!\!\!NIP},~~~~~~i,j=1,2.
\end{equation}
It is introduced by \citet{Znak2023} in this form, connecting the rank with the spatial relationship of the eigenwavefronts.
For an edge, the eigenwavefronts share a common space curve where they osculate (see Figure \ref{fig:eigenwaves}). Notably, its projection onto the acquisition plane by NIP-wave rays gives the zero-offset focusing curve. Equations \ref{mateqs} and \ref{mateqr} coincide describing its tangent, ${\bf f}$:
\begin{equation}\label{fczo}
\left({\bf M}^{(x)}_{N}-{\bf M}^{(x)}_{NIP}\right)\!{\bf f}={\bf 0},
\end{equation}
where ${\bf M}^{(x)}_{N}$ and ${\bf M}^{(x)}_{NIP}$ are matrices of traveltime derivatives from equation \ref{eq18}.
The vector field ${\bf f}$ determines the unique azimuths on the acquisition plane at which the edge diffraction resembles 2-D diffraction. More specifically, the principal curvatures of the midpoint-offset surface become equal. The identification matrix can be transformed into wavefront-orthonormal coordinates, which are local Cartesian coordinates in the wavefront-tangent plane common to both wavefronts. In these coordinates, the wavefront curvature matrices replace the traveltime derivatives: ${\bf D}^{(w.o.)}=v_1^{-1}({\bf K}_{N}-{\bf K}_{NIP})$. Here, $v_1$ refers to the phase velocity at the midpoint, in the direction of the zero-offset propagation. The wavefront curvature matrix, ${\bf K}$, represents the second fundamental form of a wavefront surface, provided that the surface is parameterized with the wavefront-orthonormal coordinates. The ranks of ${\bf D}^{(w.o.)}$ and ${\bf D}$ are equal. The homogeneous linear system with matrix ${\bf D}^{(w.o.)}$ gives the direction of osculation on the wavefront-tangent plane. In this direction, the curvatures of the wavefront sections coincide.

For finite offsets, there is also a criterion based on the mutual arrangement of wavefronts.
Imagine an extra point diffractor at the point of edge diffraction for a particular source-receiver pair, ${\bf x}_d$. The imaginary and actual diffracted wavefronts osculate, sharing a common family of rays that form the receiver side of the diffraction cone. The slowness vectors along the receiver-side focusing curve also coincide. This is illustrated in equation \ref{eq5_a}. Differentiating it now with respect to $s_2$ yields an alternative system of equations for the focusing curve:
\begin{equation}\label{mateqr2}
{\bf D}'{\bf f}_r={\bf 0},
\end{equation}
\begin{equation*}
D'_{ij}=2\!\left(\!\frac{\partial^2 t}{\partial x_i^{\!(2)}\!\partial x_j^{\!(2)}}({\bf x}_s;{\bf x}_r)-\frac{\partial^2 \tau}{\partial x_i\partial x_j}({\bf x}_r;{\bf x}_d)\!\right)\!,~i,j=1,2.
\end{equation*}
Note that the matrix ${\bf D}'$ is symmetric. In the case of edge diffractions, it has rank one. Assumption of no caustics for these two wavefronts allows us to introduce the second-order derivatives. Moreover, it guarantees regularity of the focusing curve \citep[see Appendix A in][]{Znak2023}, which justifies its differentiation with respect to the arclength.
The focusing curve tangent and the osculation direction on the wavefront-tangent plane are related as for the eigenwavefronts. ${\bf D}'={\bf 0}$ when the actual wave is a point diffraction itself. Unfortunately, the data do not contain time derivatives of the imaginary point diffraction. In a Cartesian coordinate system aligned with the focusing curve, three of four second-order derivatives of the imaginary point diffraction equal the derivatives of edge diffraction, so they can be estimated. However, the one defined orthogonally to the focusing curve remains unknown. This excludes the event identification with ${\bf D}'$.

Deriving the NIP-wave theorem, we find a representation for the matrices ${\bf D}$ and ${\bf D}’$  (see equations \ref{long_formulas}):
\begin{align}\label{decomposition_1}
&D_{ij}=2\frac{\partial s_d}{\partial x_i^{\!(1)}}({\bf x}_s; {\bf x}_r)\! \sum\limits_{k=1}^{3}\!\frac{\partial^2 \tau }{\partial x_j \partial y_k}({\bf x}_r;{\bf x}_d)\frac{dx^e_k}{ds}(s_d),\nonumber\\
&D_{ij}'=2\frac{\partial s_d}{\partial x_i^{\!(2)}} ({\bf x}_s; {\bf x}_r)\!\sum\limits_{k=1}^{3}\!\frac{\partial^2 \tau }{\partial x_j \partial y_k}({\bf x}_r;{\bf x}_d)\frac{dx^e_k}{ds}(s_d),\nonumber\\
&~~~~~~~~~~~~~~~~~~~~~~~~~~~~~~~~~~~~~~~~~~~~~~~~~~~~i,j=1,2.\end{align}
Here, $x_k^e(s)$ are components of the edge curve that is parameterized with its arc length. A symmetric function $s_d({\bf x}_s;{\bf x}_r)$ determines the parameter value at which the diffraction occurs.
Representation \ref{decomposition_1} shows that ${\bf D}'={\bf D}$ at zero offset. The presence of such a decomposition, in the form $A_{ij}=b_ic_j$, automatically guarantees that $\mbox{rank}{\bf D}=\mbox{rank}{\bf D'}=1$, unless none of the involved vectors vanishes. Let us show they do not.
${\bf D}'\ne{\bf 0}$ for an edge diffraction. Otherwise, its wavefront curvature matrix would coincide with the wavefront curvature matrix of the imaginary point diffraction. It can be uniquely backpropagated into the subsurface using dynamic ray tracing. Close to the diffractor, this would lead to the unlimited growth of both principal curvatures of the wavefront, which does not occur for linear diffractors. 
Therefore, neither $\nabla_{\!{\bf x}_r} s_d$ nor the second vector in the decomposition of ${\bf D}$ and ${\bf D}’$  vanish. Swapping the source and receiver additionally gives $\nabla_{\!{\bf x}_s} s_d\ne{\bf 0}$. This way, none of the vectors in the decomposition of ${\bf D}$ vanishes and, therefore, ${\bf D}\ne0$.

Using expressions \ref{eq10} or \ref{tsr}, the finite-offset wave type characteristic can be directly verified for a homogeneous medium. Namely, we obtain the rank-one identification matrix by differentiating any of them: 
\begin{equation}\label{D_hom}
{\bf D}=-2\frac{\cos^2\!\beta}{v^4t^3} {\bf d}_s {\bf d}_r^T,
\end{equation}
\begin{equation}\label{d_vectors}
{\bf d}_i=\left(
\setlength\arraycolsep{1pt}
\def\arraystretch{1.4}
\begin{array}{c}
2\sigma_i  h_y\displaystyle\frac{x_i}{r_i}\\
r_s + r_r+2\sigma_i \sin^2\!\beta h_y \displaystyle\frac{{\hat y}_i}{r_i}
\end{array}
\right)\!,~~~i=s,r,
\end{equation}
where \( {\hat y}_i=y_i-y_{out}\), $\sigma_s=1$, $\sigma_r=-1$.
The vectors ${\bf{d}}_s$ and ${\bf{d}}_r$ correspond to the edge orientation depicted in Figure \ref{fig:cone}.
It should be noted, however, that equation \ref{D_hom} applies for any edge orientation as long as the vector components are appropriately transformed.
In addition, the rank-one matrix for the alternative criterion can be obtained directly by differentiating equation \ref{eq10} along with the traveltime of the spherical wave from the point of diffraction:
\begin{equation}
{\bf D}'=-2\frac{\cos^2\!\beta}{v^4t^3}\frac{r_s}{r_r} {\bf d}_r {\bf d}_r^T.
\end{equation}

\section{Focusing curves}
Diffractions largely demonstrate one-way propagation and focus when time-reversed in the correct model. To better constrain velocity or other material parameters in tomographic inversion, a focusing condition can be applied to groups of rays, or each group can be attributed to a single subsurface element, which reduces the number of unknowns. However, there is a distinction between point and edge diffractions. For a 3-D point or a 2-D diffractor, all rays converge at a single point. For a 3-D edge diffractor, however,  a selection of receivers is necessary to form groups that independently focus at distinct points along the edge. 

\citet{Znak2023} organize such groupings in the post-stack domain, i.e., for zero-offset sections. This is done by picking receivers from the vicinity of the zero-offset focusing curves. These curves are built on the acquisition plane by numerical solution of the system of ordinary differential equations \ref{fczo} with matrices ${\bf M}_{N}^{(x)}$ and ${\bf M}_{NIP}^{(x)}$ extracted from data.  

This approach naturally extends to finite offsets. Rays from two focusing curves meet at the diffraction point on the edge (see Figures \ref{fig:cone} and \ref{fig:inhomogeneous_cone}). These curves can be built using equations \ref{mateqs} and \ref{mateqr} with data-derived components of matrix ${\bf D}$. For a given source location, a unit vector field on the acquisition plane is found by solving the algebraic system \ref{mateqr}. Since $\mbox{rank}{\bf D}=1$, it defines directions at different receiver locations. The receiver-side focusing curves are then the integral curves of this vector field.

Let us dwell on equations \ref{mateqr} and \ref{mateqr2}, analyzing them
from the perspective of decompositions \ref{decomposition_1}. These systems can be satisfied with the tangent to the receiver-side focusing curve due to the following identity:
\begin{equation}
\sum\limits_{j=1}^{2}\sum\limits_{k=1}^{3}\!\frac{\partial^2 \tau }{\partial x_j \partial y_k}({\bf x}_r;{\bf x}_d)\frac{dx^e_k}{ds}(s_d)\frac{dx_j}{ds_2}^{\!\!(2)}\!\!\!\!(0)=0.
\end{equation}
This identity reflects the fact that moving the ray end-point along the focusing curve does not change the edge-tangent component of the slowness vector in depth. If we substitute equation \ref{additional_expression} into decompositions \ref{decomposition_1}, we get another way to represent  the identification matrices:
\begin{align} \!\!{\bf D}&=-2F_{ss} \nabla_{\!{\bf x}_s} s_d\,\nabla_{\!{\bf x}_r}^T s_d,\nonumber\\[0.5em] \!\!{\bf D}'\!&=-2F_{ss} \nabla_{\!{\bf x}_r} s_d\,\nabla_{\!{\bf x}_r}^T s_d, \end{align} where the coefficient $F_{ss}$ is the second derivative of the Fermat function with respect to the edge parameter.
In light of this decomposition, equations \ref{mateqr} and \ref{mateqr2} hold because moving the receiver along the focusing curve does not affect the diffraction point: 
\begin{equation} \frac{ds_d}{ds_2}=\sum\limits_{j=1}^2\frac{\partial s_d}{\partial x_j^{\!(2)}}\frac{dx_j}{ds_2}^{\!\!(2)}=0. \end{equation}
From this, and from a similar equation for the opposite branch, we conclude that the gradients that make up the decompositions are orthogonal to the corresponding focusing curves:
\begin{equation}
\nabla_{\!{\bf x}_s}^T {\large s_d}\,{\bf f}_s=0, ~~~\nabla_{\!{\bf x}_r}^T {\large s_d}\,{\bf f}_r=0.
\end{equation}
The focusing curves are contour lines of $s_d({\bf x}_s;{\bf x}_r)$, with one of the arguments fixed.

Finally, to show how equations \ref{mateqs} and \ref{mateqr} apply to the triple-square-root moveout, we need the expressions for the tangents to the focusing curves. For a homogeneous medium, we can find these curves explicitly using the formula for the diffraction point \ref{eq9}. We do this by setting the expression for $y_d$ (with fixed source and receiver) equal to the same expression, but with $y_r$ replaced by $y$ and $r_r$ replaced by $(x^2+\sin^2\!\beta(y-y_{out})^2)^{1/2}$.
After some transformations, we find that the curves represent branches of hyperbolas:
\begin{equation}\label{eq21}
\frac{(y-y_{out}-y_0)^2}{a_y^2}-\frac{x^2}{a_x^2}=1
\end{equation}
with semi-axes 
\begin{align}
&a_x^2=\frac{\sin^2\!\beta(r_r {\hat y}_s+r_s {\hat y}_r)^2}{(r_s+r_r)^2-4\sin^2\!\beta h_y^2},\nonumber\\
&a_y^2=\frac{4\sin^2\!\beta h_y^2(r_r  {\hat y}_s+r_s {\hat y}_r)^2}{((r_s+r_r)^2-4\sin^2\!\beta h_y^2)^2}
\end{align}
and a shift parameter
\begin{equation}
y_0=\frac{(r_s+r_r)(r_r {\hat y}_s+r_s{\hat y}_r)}{(r_s+r_r)^2-4\sin^2\!\beta h_y^2}.
\end{equation}
Their tangent vectors are
\begin{equation}\label{eq22}
{\bf f}_i=
\left(
\setlength\arraycolsep{1pt}
\def\arraystretch{2}
\begin{array}{c}
\displaystyle
\!\!\left(1-\frac{4\sin^2\!\beta h_y^2}{(r_s+r_r)^2}\right)\! {\hat y}_i-\frac{r_r {\hat y}_s+r_s {\hat y}_r}{r_s+r_r}\\
\displaystyle
\frac{4h_y^2 x_i}{(r_s+r_r)^2} 
\end{array}
\right)\!,~~~i=s,r.
\end{equation}
The vectors \ref{d_vectors}
are orthogonal to the focusing curves:
${\bf d}_s^T{\bf f}_s=0$ and ${\bf d}_r^T {\bf f}_r=0$. As a consequence,  equations \ref{mateqs} and \ref{mateqr} are fulfilled. This is not a surprise since 
as we can make sure differentiating the explicit formula \ref{eq9}, they are proportional to the gradients:
\begin{equation}
\nabla_{\!{\bf x}_s} {\large y_d}=\frac{\cos^2\!\!\beta \, r_r}{(r_s+r_r)^2}{\bf d}_s ,~~~\nabla_{\!{\bf x}_r} {\large y_d}=\frac{\cos^2\!\!\beta \,r_s}{(r_s+r_r)^2}{\bf d}_r.
\end{equation}
At zero offset, ${\bf d}_s={\bf d}_r$ and $\nabla_{\!{\bf x}_s} {\large y_d}=\nabla_{\!{\bf x}_r} {\large y_d}$. These vectors align with the edge. The branches of hyperbolas merge to straight lines that are orthogonal to the edge.

\section{Discussion}
To organize the 3-D wavefield classification, a tool is needed that automatically extracts the diffraction identification matrix from the data. The matrix components represent traveltime derivatives and can be estimated by coherence analysis. In this framework, the local traveltimes are parameterized by their derivatives. Parameter values that maximize a coherence measure provide estimates of the wavefront attributes. Depending on the context, "wavefront attributes" can refer to either these estimates or the physical quantities themselves. A technique known as the common-reflection-surface (CRS) stack achieves this for both zero-offset \citep{Jaeger2001} and finite-offset \citep{Zhang2001} wavefront attributes.

The CRS stack is usually formulated in the midpoint-offset domain, and the identification matrix can be formulated accordingly. However, its components, being mixed source-receiver derivatives, can be estimated directly as attributes in the source-receiver domain. For this purpose, a formulation of local stacking in the source-receiver domain may be utilized, like one presented by \citet{Bakulin2020}.

Although extending the CRS stack to 3-D is theoretically straightforward, computational expenses increase substantially. First, the dimensionality of the search space, i.e., the number of wavefront attributes, increases. In 2-D, the zero-offset version has three attributes (one slowness and two second-order derivatives), and five are present in the finite-offset version (two slownesses and three 
second-order derivatives). In 3-D, it is already eight for the zero-offset case (two slowness vector components and two symmetric $2\times2$ matrices), while the finite-offset one needs fourteen (two components for each of two slowness vectors, two symmetric and one non-symmetric matrices). Secondly, the source-receiver space is 4-D. Finally, the attribute volumes become higher-dimensional.  Namely, the 2-D zero-offset sections turn into 3-D zero-offset volumes. In the case of finite offsets, for each source position, 3-D volumes of wavefront attributes need to be retrieved, resulting in a 5-D total volume. \citet{Znak2023} have demonstrated the 3-D event classification using the zero-offset version, with the SEG/EAGE Salt model dataset as an example. However, further progress requires acceleration. As one possibility, machine learning methods for rapid wavefront attribute extraction are being developed \citep{Gadylshin2023}. 

In addition to edge diffractions, other types of bulk waves exist, at least those arriving as bulk waves, that also have a rank-one matrix of mixed source-receiver traveltime derivatives. These are head waves and waves diffracted by smooth bodies, as described by \citep{Keller1962}. For these waves, matrix degeneration occurs due to the presence of curves on the acquisition surface, similar to the focusing curves, which select rays following the same path along the scattering body. Both wave types appear at larger offsets: head waves require critical incidence, whereas surface-diffracted waves need an even more unusual configuration involving grazing incidence. In seismograms, head waves show up as fast arrivals with small curvature, making them unlikely to be confused with diffractions. Diffractions by smooth bodies are of particular interest in global-scale seismology, as they are generated at the Earth's core-mantle boundary and provide additional information on the low-shear-velocity provinces. In principle, by solving systems analogous to \ref{mateqs} and \ref{mateqr}, one can construct the curves mentioned above. The directional patterns for these wave types should differ from those of the focusing curves. For edge diffractions, there are offset directions along the edge to which the curves are orthogonal, and also offset directions orthogonal to the edge to which the curves are parallel. For the surface-related waves, the curves are always parallel to the offset, and the source-side and receiver-side curves are separated by the region of non-existence.

Different traveltime approximations can be utilized to reduce noise, estimate physically meaningful parameters, or enhance the intensity of preferred waveforms. Performing coherence analysis and stacking with an approximation that incorporates unique properties of a specific wave type may be advantageous. This strategy has been previously implemented under the assumption of point diffractors \citep{Faccipieri2016}. We believe the new triple-square-root moveout is of interest in this regard, representing a potentially better approximation for edge diffraction than the hyperbolic one, at least in areas of mild wave velocity variations.

\section{Conclusions}
We have identified finite-offset properties of 3-D edge diffractions. Events of this type are routinely registered in applied seismics and by ground-penetrating radar. The local structure of the traveltime in 4-D source-receiver space allows an unambiguous discrimination of edge diffractions, point diffractions, and reflections. This structure can be analyzed using wavefront attributes accessible from the data. The wavefront attributes also enable a decomposition of edge diffractions into independent patches, each focusing at a different point along the edge. Both observations are general and apply to arbitrary curved linear diffractors in inhomogeneous and anisotropic media.  However, to illustrate this with a specific example, we have derived an exact solution — the triple-square-root moveout — that describes traveltime from an arbitrarily oriented edge in homogeneous and isotropic media. This solution matches numerical simulations and demonstrates the predicted peculiarities. As the technology of wavefront attribute extraction develops, 3-D wave processing can benefit from these findings.

\section{Acknowledgments}
This work was funded by a research fellowship from the University of Hamburg by contributions of the Wave Inversion Technology (WIT) consortium.

\appendices

\section{Singularity of the two-way \\geometrical spreading}
  \label{ap:0}
The square of geometrical spreading $L$ defines the area of the cross-section of an infinitesimal ray tube as $dS=L^2 d\gamma_1 d\gamma_2$, where $\gamma_1$ and $\gamma_2$ denote the ray coordinates.

There is a link between the two-way geometrical spreading of reflections and the matrix of mixed second-order traveltime derivatives. First, \citet{Hubral1983} showed that the spreading of zero-offset reflections can be estimated using the data-derived traveltime derivatives of the eigenwavefronts.
Specifically, its square is inversely proportional to the determinant of the difference:
\begin{equation}\label{gs1}
L^2\propto\frac{1}{\mbox{det}\!\left({\bf M}_{N}^{(x)}-{\bf M}_{NIP}^{(x)}\right)}.
\end{equation}
At about the same time, \citet{Gritsenko1984} suggested using the mixed source-receiver traveltime derivatives to determine the geometrical spreading of reflections. \citet{Goldin1986} also discussed this approach. 
For a general finite-offset configuration, the inverse proportionality regards the diffraction identification matrix:
\begin{equation}\label{gs2}
L^2\propto\frac{1}{\mbox{det}{\bf D}}.
\end{equation}
This relation turns into \ref{gs1} in the zero-offset case. One can find its precise version in the book of \citet[][  see equation 4.10.50]{Cerveny2001}.

We emphasize these relations because they enable us to look at the wave-type characteristic from a different perspective. Regarding kinematics, point and edge diffraction can both be viewed as extreme cases of reflection in the limit of infinite curvature of the reflecting boundary. 
On the other hand, during the diffraction process, an infinitely narrow ray tube unfolds into a finite ball for point diffractions. For edge diffractions, it transforms into an infinitely thin layer between two cones. The cross-section of the diffracted ray tube is not twice infinitesimal. As a result, the two-way geometrical spreading of diffractions approaches infinity, and the denominators of equations \ref{gs1} and \ref{gs2} vanish. This indicates degeneration of the identification matrix.

\section{NIP-wave theorem for 3-D diffractions}
  \label{ap:1}
Consider a midpoint, ${\bf x}_m$, and a zero-offset ray that connects it to a point of a scatterer in depth (Figure \ref{fig:eigenwaves}). By Fermat's principle, the zero-offset incidence is normal. In an anisotropic medium, although the ray becomes inclined relative to the scatterer, its slowness vector remains normal. Therefore, this depth point is called the normal-incidence-point (NIP). An imaginary wave emanating from the NIP is referred to as a NIP-wave. Different midpoints produce a family of such wavefronts with traveltimes on the acquisition plane $t^{NIP}({\bf x};{\bf x}_m)$. The second argument specifies a wavefront from the set.

The NIP-wave theorem locally describes the two-way traveltimes $t({\bf x}_m; {\bf h})$ in terms of the NIP-waves. Specifically, it states that
\begin{equation}\label{theorem_appendix}
\frac{\partial^2 t}{\partial h_i \partial h_j}({\bf x}_m;{\bf 0})=2 {\frac{\partial^2 t}{\partial x_i \partial x_j}}^{\!\!\!\!\!\!\!NIP}\!\!\!\!\!({\bf x}_m;{\bf x}_m),~~~i,j=1,2.
\end{equation}
Equivalently, up to the second order, the two-way time is a sum of the times of two branches connecting the NIP to the source and receiver, as if the scattering point doesn't migrate depending on the offset. In other words, the two-way traveltime is approximated by that of the point diffraction.

The right-hand side of equation \ref{theorem_appendix} contains the second-order wavefront derivatives. Therefore, it is necessary that this wavefront does not hit a caustic at the point of consideration. Therefore, a first necessary assumption is that there is no caustic of the NIP-wave at the midpoint.

For point diffractions, the theorem holds trivially.
The two-point traveltime for a diffractor at ${\bf x}_d$ reads
\begin{equation}
t({\bf x}_s; {\bf x}_r)=\tau({\bf x}_s;{\bf x}_d)+\tau({\bf x}_r;{\bf x}_d).
\end{equation}
In the midpoint-offset representation,
\begin{equation}\label{pd_time}
\tilde{t}({\bf x}_m; {\bf h})=\tau({\bf x}_m-{\bf h};{\bf x}_d)+\tau({\bf x}_m+{\bf h};{\bf x}_d).
\end{equation}
Differentiation of equation \ref{pd_time} with respect to the offset components at ${\bf h}={\bf 0}$ yields
\begin{equation}\label{theorem_point}
\frac{\partial^2 t}{\partial h_i \partial h_j}({\bf x}_m;{\bf 0})=2 {\frac{\partial^2 \tau}{\partial x_i \partial x_j}}({\bf x}_m;{\bf x}_d),~~~i,j=1,2.
\end{equation}
When considered for different midpoints, the zero-offset rays arrive at the same point in depth, namely the point diffractor, which can be formally referred to as NIP. All such NIP-wavefronts coincide.

For edge diffractions, we follow an approach originally proposed for 2-D reflections by \citet{Chernjak1979}. In English, it is described in the paper by \citet{Fomel2001}.

Firstly, we express the derivatives with respect to the offset in terms of the derivatives of the two-point traveltime. From the identity $\tilde t({\bf x_m};{\bf h})=t({\bf x}_s({\bf x}_m;{\bf h});{\bf x}_r({\bf x}_m;{\bf h}))$, it follows that
\begin{align}
\frac{\partial^2 \tilde t}{\partial h_i \partial h_j}=
\frac{\partial^2 t}{\partial x_i^{\!(1)} \!\partial x_j^{\!(1)}}
+\frac{\partial^2 t}{\partial x_i^{\!(2)} \!\partial x_j^{\!(2)}}
-&\frac{\partial^2 t}{\partial x_i^{\!(1)} \!\partial x_j^{\!(2)}}
-\frac{\partial^2 t}{\partial x_j^{\!(1)} \!\partial x_i^{\!(2)}},\nonumber\\
&~~~i,j=1,2.
\end{align}
At zero offset, the time reciprocity leads to identities 
\begin{align}
&\frac{\partial^2 t}{\partial x_i^{\!(1)} \!\partial x_j^{\!(1)}}=\frac{\partial^2 t}{\partial x_i^{\!(2)}\! \partial x_j^{\!(2)}},~~~~\frac{\partial^2 t}{\partial x_i^{\!(1)}\! \partial x_j^{\!(2)}}=\frac{\partial^2 t}{\partial x_j^{\!(1)}\! \partial x_i^{\!(2)}},\nonumber\\
&~~~~~~~~~~~~~~~~~~~~~~~~~~~~~~~~~~~~~~~~~~~~~~~~~~~~i,j=1,2,
\end{align}
and, consequently, to the expression:
\begin{align}\label{difference}
&\frac{\partial^2 \tilde t}{\partial h_i \partial h_j}=2\left(
\frac{\partial^2 t}{\partial x_i^{\!(2)} \!\partial x_j^{\!(2)}}({\bf x}_m;{\bf x}_m)
-\frac{\partial^2 t}{\partial x_i^{\!(1)} \!\partial x_j^{\!(2)}}({\bf x}_m;{\bf x}_m)\right)\nonumber\\
&~~~~~~~~~~~~~~~~~~~~~~~~~~~~~~~~~~~~~~~~~~~~~~~~~~~~i,j=1,2.
\end{align}

Consider a function to be optimized according to Fermat's principle:
\begin{equation}\label{Fermat}
F({\bf x}_s; {\bf x}_r;s)=\tau({\bf x}_s;{\bf x}_e(s))+\tau({\bf x}_r;{\bf x}_e(s)),
\end{equation}
where ${\bf x}_e(s)$ stands for the edge curve parameterized with its arc length $s$. 
The actual diffraction point, with $s=s_d$, makes the function stationary: 
\begin{align}\label{Fermat's_equation}
&F_s=\frac{\partial F}{\partial s}({\bf x}_s;{\bf x}_r;s_d)=\\
&~~~~\sum\limits_{i=1}^{3}\left(\frac{\partial \tau}{\partial y_i}({\bf x}_s;{\bf x}_e(s_d))+\frac{\partial \tau}{\partial y_i}({\bf x}_r;{\bf x}_e(s_d))\right)\!\frac{dx_i^e}{d s}(s_d)=0\nonumber.
\end{align}
Substituting $s_d$ into equation \ref{Fermat} gives the actual traveltime of the diffracted wave. Equation \ref{Fermat's_equation} shows that the edge-tangent component of the slowness vector is conserved. This is Keller's law. In particular, when ${\bf x}_s={\bf x}_r$, the slowness vector becomes orthogonal to the edge. 

Additionally, we require non-vanishing of the second derivative:
\begin{equation}\label{condition}
F_{ss}=\frac{\partial^2\!F}{\partial s^2}({\bf x}_s;{\bf x}_r;s_d)\ne0.
\end{equation}
When this condition holds, the Fermat function achieves an extremum. By the implicit function theorem, there exists a function $s_d({\bf x}_s;{\bf x}_r)$ that describes the diffraction point location. Notably, we have derived this function in explicit form for a straight edge in a homogeneous medium (see equation \ref{eq9}). In the final part of the section, we explain the physical meaning of condition \ref{condition} at zero offset, i.e., $F_{ss}({\bf x}_m; {\bf x}_m;s_d)\ne0$. This ensures the implicit function is defined for ${\bf x}_s$ and ${\bf x}_r$ in the vicinity of ${\bf x}_m$, which suffices for the NIP-wave theorem derivation.

The  two-point two-way traveltime reads
\begin{align}\label{Fermat_substituted}
t({\bf x}_s;{\bf x}_r)&=F({\bf x}_s; {\bf x}_r;s_d({\bf x}_s; {\bf x}_r))=\\
&\tau({\bf x}_s;{\bf x}_e(s_d({\bf x}_s;{\bf x}_r)))+\tau({\bf x}_r;{\bf x}_e(s_d({\bf x}_s;{\bf x}_r)))\nonumber.
\end{align}
Since ${\bf x}_s$ and ${\bf x}_r$ are symmetrically present in equation \ref{Fermat's_equation}, its solution 
and consequently the traveltime become reciprocal: $s_d({\bf x}_s;{\bf x}_r)=s_d({\bf x}_r;{\bf x}_s)$ and $t({\bf x}_s;{\bf x}_r)=t({\bf x}_r;{\bf x}_s)$.
Differentiating the traveltime \ref{Fermat_substituted} with respect to the receiver position yields
\begin{align}
\frac{\partial t}{\partial x_i^{\!(2)}}({\bf x}_s&; {\bf x}_r)=\frac{\partial \tau}{\partial x_i}({\bf x}_r; {\bf x}_e(s_d({\bf x}_s;{\bf x}_r)))\\
&+F_s({\bf x}_s; {\bf x}_r; s_d({\bf x}_s;{\bf x}_r))\frac{\partial s_d}{\partial x_i^{\!(2)}}({\bf x}_s; {\bf x}_r),~~~i=1,2.\nonumber
\end{align}
The first factor of the second term here identically equals zero due to equation \ref{Fermat's_equation}, so
\begin{equation}
\frac{\partial t}{\partial x_i^{\!(2)}}({\bf x}_s; {\bf x}_r)=\frac{\partial \tau}{\partial x_i}({\bf x}_r; {\bf x}_e(s_d({\bf x}_s;{\bf x}_r))),~~~i=1,2.
\end{equation}
We differentiate this result again, first with respect to the source coordinates and second with respect to the coordinates of the receiver:
\begin{align}\label{long_formulas}
\frac{\partial^2 t}{\partial x_j^{\!(1)} \!\partial x_i^{\!(2)}}&=\frac{\partial s_d}{\partial x_j^{\!(1)}}({\bf x}_s;{\bf x}_r)\sum\limits_{k=1}^3\frac{\partial^2 \tau}{\partial x_i \partial y_k}({\bf x}_r;{\bf x}_d)\frac{dx_k^e}{ds}(s_d), \nonumber\\
\frac{\partial^2 t}{\partial x_i^{\!(2)} \!\partial x_j^{\!(2)}}&=\frac{\partial^2 \tau}{\partial x_i \partial x_j}({\bf x}_r;{\bf x}_d)\nonumber\\
&+\frac{\partial s_d}{\partial x_j^{\!(2)}}({\bf x}_s;{\bf x}_r)\sum\limits_{k=1}^3\frac{\partial^2 \tau}{\partial x_i \partial y_k}({\bf x}_r;{\bf x}_d)\frac{dx_k^e}{ds}(s_d),\nonumber\\
~~~~~~~~&~~~~~~~~~~~~~~~~~~~~~~~~~~~~~~~~~~i,j=1,2,
\end{align}
where ${\bf x}_d={\bf x}_e(s_d({\bf x}_s;{\bf x}_r))$. 

The first identity specifies the entries of the identification matrix ${\bf D}$. 
The second identity enables the determination of the components of another important matrix, denoted as ${\bf D'}$. In the main text, an additional observation is utilized: the vector with index $i$ on the right-hand side of equations \ref{long_formulas} is proportional to the gradient of the implicit function:
\begin{align}\label{additional_expression}
\sum\limits_{k=1}^3\frac{\partial^2 \tau}{\partial x_i \partial y_k}({\bf x}_r;&{\bf x}_d)\frac{dx_k^e}{ds}(s_d)=\\
&~-F_{ss}({\bf x}_s;{\bf x}_r;s_d)\frac{\partial s_d}{\partial x_i^{(2)}}({\bf x}_s;{\bf x}_r), ~~~i=1,2.\nonumber
\end{align}
This can be done by differentiation of equation \ref{Fermat's_equation} with respect to ${\bf x}_r$ after substitution of $s_d({\bf x}_s;{\bf x}_r)$.

Because $s_d({\bf x}_s;{\bf x}_r)$ is symmetric,  $\partial s_d/{\partial x_1^{(2)}}=\partial s_d/\partial x_1^{(1)}$ and $\partial s_d/{\partial x_2^{(2)}}=\partial s_d/\partial x_2^{(1)}$  when ${\bf x}_s={\bf x}_r$. The terms comprising these derivatives cancel out in the difference \ref{difference} computed with the help of equations \ref{long_formulas}, and 
\begin{align}
&\frac{\partial^2 t}{\partial h_i \partial h_j}({\bf x}_m;{\bf 0})=2\frac{\partial^2 \tau}{\partial x_i \partial x_j}({\bf x}_m;{\bf x}_e(s_d({\bf x}_m;{\bf x}_m))),\nonumber\\
&~~~~~~~~~~~~~~~~~~~~~~~~~~~~~~~~~~~~~~~~~~~~~~~~~~i,j=1,2,
\end{align}
which proves the NIP-wave theorem.

Equations \ref{long_formulas} include mixed derivatives of the direct traveltime $\tau$ with respect to the acquisition plane point and the depth point. These derivatives can be determined as long as the NIP-wave has no caustic at the midpoint \citep[section 4.9.4 in][]{Cerveny2001}. Therefore, no additional conditions are required in this regard.

The assumption \ref{condition}, though,
imposes a restriction on the edge line, which can be interpreted in terms of the N-wave in the zero-offset case. For zero offset,
\begin{align}\label{Fss}
F_{ss}({\bf x}_m;{\bf x}_m;s)=2&\sum\limits_{i,j=1}^3\frac{\partial^2\tau}{\partial y_i \partial y_j}({\bf x}_m;{\bf x}_e(s))\frac{dx^e_i}{ds}\frac{dx^e_j}{ds}\nonumber\\
&+2\sum\limits_{i=1}^3\frac{\partial \tau}{\partial y_i}({\bf x}_m;{\bf x}_e(s))\frac{d^2x^e_i}{ds^2}.
\end{align}
Let $s=s_d$, i.e., it makes the Fermat function stationary. Therefore, it determines a point of normal incidence. The second-order derivatives of the incident front in equation \ref{Fss} are then justified due to the reciprocal relation for the geometrical spreading. If the NIP-wave has no caustic at the midpoint, neither does the incident wave at NIP. Denote the edge-tangent unit vector as ${\bf t}_e$, ${\bf t}_e = d{\bf x}_e/ds$. The second derivative $d^2{\bf x}_e/ds^2 = K_e{\bf n}_e$, where $K_e$ is the edge curvature and ${\bf n}_e$ is the normal to the edge. Then,
\begin{equation}
F_{ss}=2{\bf t}_e^T{\hat {\bf M}}^{(x)}{\bf t}_e+2v_0^{-1}K_e{\bf n}_w^T{\bf n}_e.
\end{equation}
$v_0$ is the phase velocity at NIP in the direction normal to the incident wavefront, ${\bf n}_w$. The hat sign here indicates a $3 \times 3$ matrix.
Since ${\bf t}_e$ is unit length and lies in the wavefront-tangent plane, ${\bf t}_e^T{ \hat {\bf M}}^{(x)}{\bf t}_e=v_0^{-1}{{\bf t}_e^{(w.o.)\,T}}{\bf K}{\bf t}_e^{(w.o.)}$,
whith ${\bf K}$ being the curvature matrix of the incident front and ${\bf t}_e^{(w.o.)}$ representing the vector ${\bf t}_e$ in the wavefront-orthonormal coordinates.
Since the curvature of the wavefront in direction ${\bf u}$ can be calculated by formula $\kappa({\bf u})={{\bf u}^{(w.o.)\,T}}{\bf K}{\bf u}^{(w.o.)}$, the condition $F_{ss}({\bf x}_m; {\bf x}_m;s_d)=0$ transforms into
\begin{equation}\label{n_wave_condition}
\kappa({\bf t}_e)=-K_e{\bf n}_w^T{\bf n}_e.
\end{equation}
The curvature of the incident wavefront in the direction of the edge equals the projection of the vector curvature of the edge onto the wavefront normal.
For example, imagine a diffractor in the shape of a circle positioned in a homogeneous medium, such that the incident spherical front passes through all its points simultaneously. For such a diffractor, the Fermat function doesn't have an extremum since all incident rays are equivalent.
The curvatures of the sphere and the nested circle are related by equation \ref{n_wave_condition}. The N-wave in this example represents an inflating torus. It converges to its axis of symmetry, which passes through the source on the acquisition plane. In this way, the N-wave has a caustic point of the first kind at the midpoint, which means the ray tube shrinks to a line there. 

Finally, we use dynamic ray tracing to demonstrate this in the general case. Consider the wavefront-orthonormal basis vectors, ${\bf e}_1(t)$ and ${\bf e}_2(t)$, defined along the zero-offset ray in the upgoing direction with $t=0$ at the NIP. The basis is conventionally introduced using a $3\times3$ system of linear differential equations 
\begin{equation}\label{basis_system}
\frac{d{\bf e}_i}{dt}={\bf e}_i^T\nabla_{\!\bf x}v({\bf x},{\bf e}_3)
\,{\bf e}_3, ~~~~~~i=1,2.
\end{equation}
The wavefront normal ${\bf e}_3$ and the gradient of the phase velocity $v$ are specified by kinematic ray tracing. For our purposes, we use special initial conditions: ${\bf e}_1(0)={\bf t}_e\times {\bf e}_3(0)$, ${\bf e}_2(0)={\bf t}_e$, where ${\bf e}_3(0)=-{\bf n}_w$.
The $4\times4$ dynamic ray tracing in wavefront-orthonormal coordinates (isotropic by \citet{Popov1978}, anisotropic by \citet{Kashtan1983, Petrashen1984, Bakker1996}) represents a linear system of eight differential equations in the block form:
\begin{equation}\label{DRT}
\frac{d}{dt}\!\left(
\setlength\arraycolsep{1.2pt}
\def\arraystretch{1}
\begin{array}{c}
{\bf Q}\\
{\bf P}
\end{array}\right)={\bf S}^{(4)}\!\left(
\setlength\arraycolsep{1.2pt}
\def\arraystretch{1}
\begin{array}{c}
{\bf Q}\\
{\bf P}
\end{array}\right),
\end{equation}
where ${\bf Q}$ and ${\bf P}$ are $2\times2$ matrices comprising projections of the ray variations:
\begin{equation}
\label{eq:proj}
Q_{ij}={\bf e}_i^T\!\frac{\partial {\bf x}}{\partial \gamma_j},~~~~P_{ij}={\bf e}_i^T\!\frac{\partial {\bf p}}{\partial \gamma_j},~~~~i,j=1,2.
\end{equation}
For the N-wave in the edge diffractor case, $\gamma_1$ is the angle $\varphi$ between the shooting wavefront normal and the edge normal. $\gamma_2$ is the arc length of the edge. The $4\times4$ matrix ${\bf S}^{(4)}$ depends on the medium, the ray, and the choice of the initial basis vectors. The ray Jacobian, responsible for the amplitude of the ray method, is determined as $J=v\,\mbox{det} \bf Q$. Away from a caustic point, $\mbox{det} {\bf Q}\ne 0$, and the wavefront curvature matrix ${\bf K}=v{\bf P}{\bf Q}^{-1}$. 
A solution to the system \ref{DRT} can be expressed through the $2\times2$ blocks of the $4\times4$ upward propagator ${\bf \Pi}$:
\begin{equation}
\setlength\arraycolsep{1.2pt}
\def\arraystretch{1}
\left(
\begin{array}{c}
{\bf Q}\\
{\bf P} 
\end{array}
\right)=
{\bf \Pi}\left(
\begin{array}{c}
{\bf Q}(0)\\
{\bf P}(0) 
\end{array}
\right),
~~~
\setlength\arraycolsep{1.2pt}
\def\arraystretch{1}
{\bf \Pi}=
\left(
\setlength\arraycolsep{1.2pt}
\def\arraystretch{1}
\begin{array}{cc}
{\bf Q}_1&{\bf Q}_2\\
{\bf P}_1&{\bf P}_2 
\end{array}
\right).
\end{equation}
The first blocky column is defined with initial conditions ${\bf Q}(0)={\bf I}$, ${\bf P}(0)={\bf 0}$ and the second one with ${\bf Q}(0)={\bf 0}$, ${\bf P}(0)={\bf I}$, where ${\bf I}$ denotes the identity matrix.
When the ray intersects layer boundaries, the propagator matrix incorporates the corresponding jumps in ${\bf Q}$ and ${\bf P}$.

For the N-wave,
\begin{equation}
{\bf Q}_N={\bf Q}_1{\bf Q}_N(0)+{\bf Q}_2{\bf P}_{\!N}(0).
\end{equation}
Let us introduce the backward propagator ${\bf \Pi}^b$ implementing the propagation from the midpoint to the NIP,  with columns ${\bf Q}_1^{b}$, ${\bf P}_1^{b}$ and ${\bf Q}_2^{b}$, ${\bf P}_2^{b}$, such that ${\bf e}_1^b=-{\bf e}_1$, ${\bf e}_2^b={\bf e}_2$, ${\bf e}_3^b=-{\bf e}_3$. Here, the equality is assumed for the same point in space. The upgoing propagator is backward to this one. Therefore, we can relate them \citep[see equation 4.4.113 in][]{Cerveny2001}:
\begin{equation}
{\bf \Pi}=\left(
\setlength\arraycolsep{1.2pt}
\def\arraystretch{1}
\begin{array}{cc}
{\overline{{\bf P}_2^{b}}}^{T}&{\overline{{\bf Q}_2^b}}^T\\
{\overline{{\bf P}_1^{b}}}^T&{\overline{{\bf Q}_1^b}}^T
\end{array}
\right)
\end{equation}
The line over a matrix negates the off-diagonal elements. Therefore,
\begin{equation}
{\bf Q}_N={\overline{{\bf P}_2^{b}}}^T\!\!{\bf Q}_N(0)+{\overline{{\bf Q}_2^b}}^T\!\!{\bf P}_{\!N}(0)=v_1\!\!\left({\overline{{\bf P}}}^T\!\!{\bf Q}_N(0)+{\overline{{\bf Q}}}^T\!{\bf P}_N(0)\right).
\end{equation}
We use ${\bf Q}$ and ${\bf P}$ to represent the incident wave, which is generated by a point source at the surface. Since for a point source ${\bf Q}(0)={\bf 0}$ and ${\bf P}(0)={\bf I}/v(0)$, ${\bf Q}={\bf Q}_2^b/v_1$ and ${\bf P}={\bf P}_2^b/v_1$. On the other hand, ${\bf Q}_{NIP}={\bf Q}_2/v_0={\overline{{\bf Q}_2^b}}^T\!\!\!/v_0$. Therefore, in the assumption of no caustic for the NIP-wave, $\mbox{det} {\bf Q}_{NIP}\ne0$, there is no caustic of the incident wave as well, $\mbox{det} {\bf Q}\ne0$, and we can factor out:
\begin{align}\label{eqb26}
{\bf Q}_N=v_1{\overline{{\bf Q}}}^T\!\!&\left(\left({\overline{{\bf P}}}\,{\overline{{\bf Q}}^{-1}}\right)^{\!T}\!\!\!{\bf Q}_N(0)+\!{\bf P}_{\!N}(0)\right)=\nonumber\\
&~~~~~~~~~~~~\frac{v_1}{v_0}{\overline{{\bf Q}}}^T\!\!\left(\,\overline{\bf K}{\bf Q}_N(0)+v_0{\bf P}_{\!N}(0)\right).
\end{align}
According to equations 4.5.60 in \citet{Cerveny2001}, initial conditions for the N-wave, a simultaneously ignited 3-D curve, read
\begin{equation}
{\bf Q}_N(0)=
\left(
\setlength\arraycolsep{1.2pt}
\def\arraystretch{1}
\begin{array}{cc}
0~&0\\
0~&1 
\end{array}
\right),~
{\bf P}_{\!N}(0)=
\frac{1}{v_0}
\left(
\setlength\arraycolsep{1.2pt}
\def\arraystretch{1}
\begin{array}{cc}
1~&T_e\\
0~&-K_e \!\cos\!\varphi 
\end{array}
\right),
\end{equation}
where $T_e$ is the torsion of the edge.
Substituting them into equation \ref{eqb26} together with $\cos\!\varphi=-{\bf n}_w^T{\bf n}_e$, we get
\begin{equation}
{\bf Q}_N=\frac{v_1}{v_0}{\overline{{\bf Q}}}^T\!\!
\left(
\setlength\arraycolsep{1.5pt}
\def\arraystretch{1}
\begin{array}{cc}
1&T_e-K_{12}\\
0 & K_{22}+K_e {\bf n}_w^T{\bf n}_e
\end{array}
\right)=\frac{v_1}{v_0}{\overline{{\bf Q}}}^T
\!\!\left(
\setlength\arraycolsep{1.5pt}
\def\arraystretch{1}
\begin{array}{cc}
1&T_e-K_{12}\\
0 & 0
\end{array}
\right)
.
\end{equation}
In the last transformation, we recall that ${\bf e}_2^b={\bf t}_e$ at NIP, which means $K_{22}=\kappa({\bf t}_e)$, and apply the condition \ref{n_wave_condition}. Thus, ${\bf Q}_N$ is a rank-one matrix at the midpoint, i.e., it describes a caustic of the first order. 

Therefore, to justify our derivation of the NIP-wave theorem, we should assume that neither of the eigenwavefronts has a caustic at the midpoint on the acquisition plane. Note, under the assumption of no caustic for the NIP-wave, a caustic of the second order, a focus point, of the N-wave is not possible for linear diffractors.

\bibliographystyle{plainnat}
\bibliography{main}

\end{document}